\documentclass[journal]{IEEEtran}
\usepackage{multirow,tabularx,caption,subcaption,bm,enumitem,kantlipsum,xcolor,hyperref,fontawesome5,balance,soul,graphicx,amsmath,arydshln,xcolor,colortbl,algorithm,algpseudocode}
 \usepackage{booktabs} 
 \usepackage{adjustbox}

\ifCLASSINFOpdf
\else
\fi

\begin{document}
%
\title{Efficient and Effective Adaptation of Multimodal Foundation Models in Sequential Recommendation}
%
%
%

\author{Junchen Fu,
        Xuri Ge*, Xin Xin, Alexandros Karatzoglou, Ioannis Arapakis, Kaiwen Zheng, Yongxin Ni,
        and Joemon M. Jose
\thanks{ 
J. Fu, K. Zheng, and J. Jose are with the School of Computing Science, University of Glasgow, Glasgow,
UK (e-mail: j.fu.3@research.gla.ac.uk, k.zheng.1@research.gla.ac.uk, and joemon.jose@glasgow.ac.uk). X. Ge is with the School of Artificial Intelligence, and X. Xin is with the School of Computer Science and Technology, both at Shandong University, Jinan, Shandong, China. (e-mail: xurigexmu@gmail.com and xinxin@sdu.edu.cn) 
A. Karatzoglou is with Amazon, Barcelona, Spain (email: alexandros.karatzoglou@gmail.com).
I. Arapakis is with Telefonica Scientific Research, Barcelona, Spain (email: arapakis.ioannis@gmail.com).
Y. Ni is with University of Science and Technology of China, Hefei, China (email: niyongxin2016@gmail.com) (\textit{* Corresponding author: Xuri Ge.)}}
}

%
%

\markboth{Journal of \LaTeX\ Class Files,~Vol.~14, No.~8, August~2015}%
{Shell \MakeLowercase{\textit{et al.}}: Bare Demo of IEEEtran.cls for IEEE Journals}
%



\maketitle

\begin{abstract}
Multimodal foundation models (MFMs) have revolutionized sequential recommender systems through advanced representation learning. While Parameter-efficient Fine-tuning (PEFT) is commonly used to adapt these models, studies often prioritize parameter efficiency, neglecting GPU memory and training speed. To address this, we introduced the IISAN framework, significantly enhancing efficiency. However, IISAN was limited to symmetrical MFMs and identical text and image encoders, preventing the use of state-of-the-art Large Language Models. To overcome this, we developed IISAN-Versa, a versatile plug-and-play architecture compatible with both symmetrical and asymmetrical MFMs. IISAN-Versa employs a Decoupled PEFT structure and utilizes both intra- and inter-modal adaptation. It effectively handles asymmetry through a simple yet effective combination of group layer-dropping and dimension transformation alignment. Our research demonstrates that IISAN-Versa effectively adapts large text encoders, and we further identify a scaling effect where larger text encoders generally perform better. IISAN-Versa also demonstrates strong versatility in our defined multimodal scenarios, which include raw titles and captions generated from images and videos. Additionally, IISAN-Versa achieved state-of-the-art performance on the MicroLens public benchmark. We release our code at \url{https://github.com/GAIR-Lab/IISAN}.
\end{abstract}


%
\IEEEpeerreviewmaketitle

\section{Introduction}\label{sec:intro}
Recent advancements in recommendation algorithms have demonstrated that utilizing powerful large language models (LLMs) and vision encoders like GPT, Llama\footnote{\url{https://www.llama.com/}, \url{https://openai.com/index/hello-gpt-4o/}}, and CLIP \cite{radford2021learning} can obtain a state-of-the-art recommendation performance \cite{yuan2023go,li2024empirical,li2023multi,hou2022learning,liu2024once}. These types of models typically combine a recommender model as the user encoder with multimodal foundation models (MFMs)\footnote{The definition of multimodal foundation models (MFMs) may vary. In this paper, we focus on separately pre-trained text and image encoders, following the approach in ~\cite{fei2022towards,radford2021learning}, where the representations of image and text are learned in an end-to-end manner. Other MFM paradigms, such as \cite{geng2023vip5,alayrac2022flamingo}, which use a large language model (LLM) as the backbone and input image encoder features into the LLM, are beyond the scope of this paper.} to encode the items. Many studies \cite{yuan2023go,li2024empirical,li2023multi} have shown that fine-tuning the item encoder can lead to optimal performance. However, these approaches are highly inefficient due to the immense computational resources required to fine-tune such large encoders.

A popular paradigm \cite{fu2024exploring} for resolving the efficiency problem includes methods such as Adapter~\cite{houlsby2019parameter} and LoRA~\cite{hu2021lora}, BitFit~\cite{zaken2021bitfit}. These approaches integrate tunable neural network modules into the backbone multimodal foundation models, collectively referred to as embedded parameter-efficient fine-tuning (EPEFT) \cite{fu2024iisan}.  While EPEFT methods have gained popularity for addressing the efficiency of trainable parameters, they still face practical efficiency challenges, such as training time and GPU memory consumption. The introduction of the IISAN~\cite{fu2024iisan}, with its decoupled structure and caching strategy for multimodal encoders, addresses these practical efficiency issues, outperforming both traditional full fine-tuning and EPEFT methods in terms of both performance and efficiency. We refer to this decoupled, parameter-efficient fine-tuning approach as DPEFT. It significantly reduces GPU memory usage by up to 15 times compared to FFT and 12 times compared to EPEFT (Adapter/LoRA). Additionally, it accelerates training time per epoch by up to 20 times compared to FFT and 16 times compared to EPEFT.

\begin{figure}
  \centering
   \includegraphics[width=\linewidth]{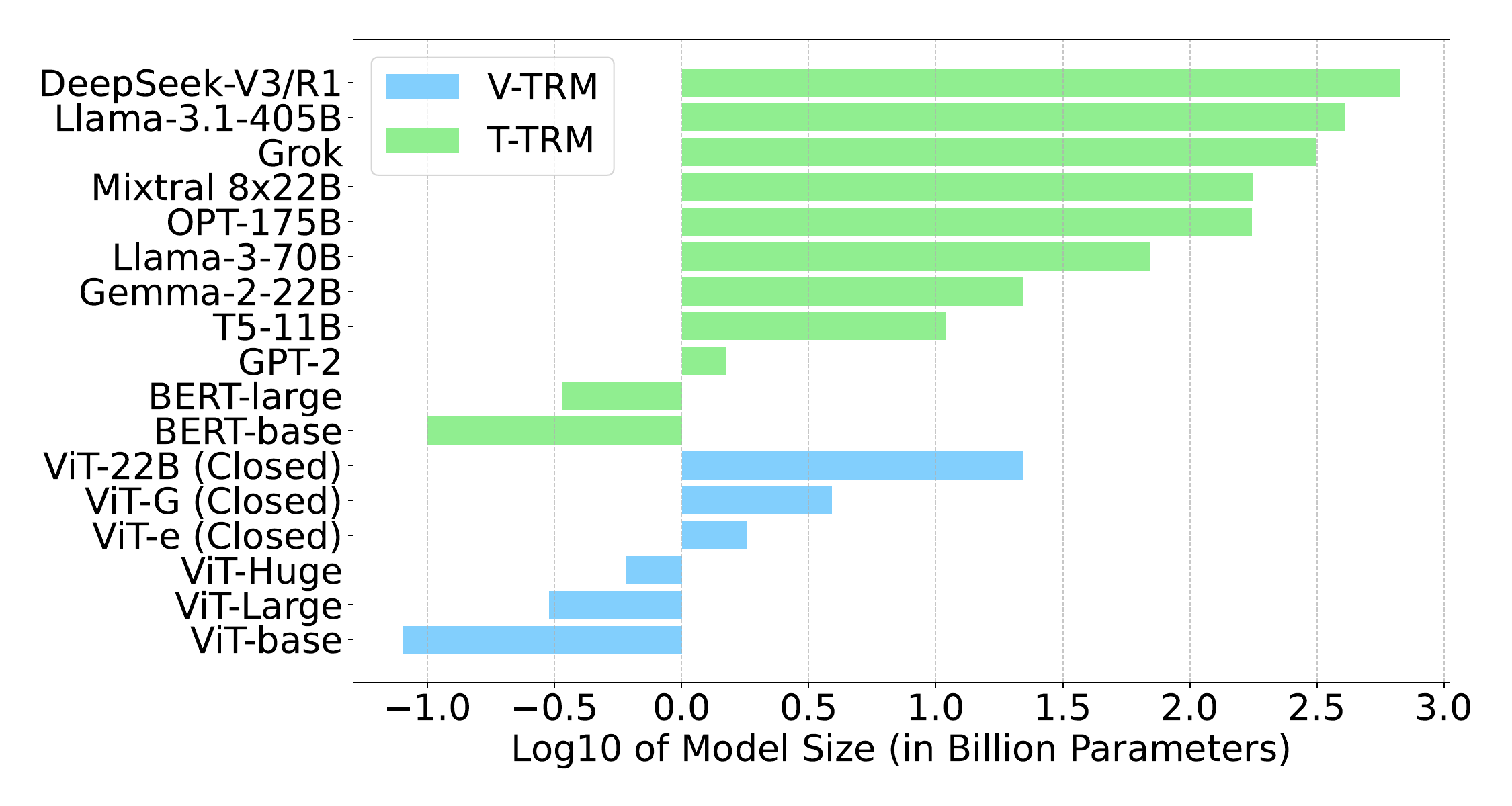}
  \caption{\textcolor{black}{Comparisons of model sizes for various popular text transformers (T-TRM) and visual transformers (V-TRM) on a logarithmic scale. Closed-source models are indicated by ``(Closed)" in the figure; all other models are open-source. }
  }
    \label{fig:compare_t_v} 
\end{figure}

 IISAN leverages both intra- and inter-modal information from multimodal foundation models by exploiting the hidden states within the layers of backbone models to achieve fine-grained information merging. Despite these advancements, merging multimodal hidden states from two different encoders presents two limitations: \textbf{(1) it can only be conducted for symmetrical multimodal transformer encoders; (2) due to the first limitation, it is hard to explore whether scaling the text encoder in IISAN with recent state-of-the-art LLMs could yield performance improvements.} These limitations do not align with the current research trend, where text encoders are often larger and more complex than visual transformers, as shown in Figure \ref{fig:compare_t_v}, which compares the state-of-the-art text and visual transformers. While vision transformers like ViT-22B \cite{dehghani2023scaling} have billions of parameters, they are all closed-source models. Therefore, the most commonly used vision transformers currently have fewer than 1 billion parameters \cite{dosovitskiy2020image}. Normally, larger pre-trained transformers generally offer better performance according to the scaling effect \cite{rosenfeld2021scaling}. To maximize the potential of pre-trained models, it is advantageous to use larger models. However, finding a vision transformer that matches the size of a text encoder (LLM) is nearly impossible due to the disparity in model sizes. Therefore, addressing the issue of asymmetrical merging is of paramount importance.

In this paper, we present an extension of the IISAN framework \cite{fu2024iisan} called the Intra- and Inter-model Side Adapted Network for Versatile Multimodal Representation (IISAN-Versa). We specifically categorize IISAN-Versa into two variants: the symmetrical IISAN-Versa (IISAN-VS), which follows the standard IISAN configuration, and the newly proposed asymmetrical IISAN-Versa (IISAN-VA), which is capable of incorporating a larger text encoder to address asymmetry between text and vision pre-trained models. For IISAN-VA, we address the challenge of asymmetric multimodal backbone networks by proposing a simple yet effective strategy that combines group layer-dropping with dimension transformation alignment. 
The IISAN-VA variant, with a larger text encoder, significantly outperforms the standard symmetrical IISAN-VS, which uses BERT-base text encoders, while maintaining high efficiency. This demonstrates IISAN-Versa's effectiveness in accommodating a larger and more powerful model. 
Moreover, we extend and validate the IISAN-Versa framework to a wider range of multimodal scenarios, particularly focusing on multimodal text, thereby validating its scalability. This extension involves the integration of diverse text forms (named multimodal text), such as titles, text captions from image covers, and video content, utilizing an open-source video recommendation dataset MicroLens~\cite{ni2023content}.

We summarize the main changes made in this study compared to our published conference version in SIGIR2024 \cite{fu2024iisan} as follows:

\begin{itemize}
\item We revise the Introduction and Related Work sections to emphasize the motivation of achieving the versatility of the new IISAN-Versa, rather than focusing only on the efficiency of IISAN. 
\item We expand the methodology section by categorizing IISAN~\cite{fu2024iisan} as the symmetrical IISAN-Versa (IISAN-VS) and introduced the asymmetrical IISAN-Versa (IISAN-VA) to incorporate LLMs as encoders.
\item We validate the effectiveness of the IISAN-VA, which incorporates state-of-the-art LLMs as text encoders, demonstrating significant performance improvements. Furthermore, we fully explore the effect of scaling laws on text encoders, observing that larger pre-trained text encoders typically lead to better performance.
 \item To gain a clearer understanding of efficiency improvement during the training process, this paper analyzes it from the perspectives of forward and backward propagation. 
\item We further validate IISAN-Versa's performance on MicroLens, a public multimodal recommendation benchmark, achieving state-of-the-art performance.
\item We also utilize pre-trained captioning models to generate multimodal texts from raw images and videos from MicroLens and validated IISAN-Versa’s versatility in this multimodal text scenario.
\end{itemize}

Our main contributions are the following:
\begin{itemize}
    \item We propose a versatile paradigm, IISAN-Versa, which adapts existing mainstream symmetric and asymmetric multimodal foundation models to enable efficient and effective multimodal sequential recommendation. It ensures the flexibility of multi-structure model adaptation and the efficiency of PEFT through minimal modifications of dimension transformation layers with LayerDrop.
    \item We identify a scaling effect in our novel asymmetric IISAN-Versa by scaling the text encoder from a smaller language model to a larger language model. This provides a new inspiration that scaling LLMs with appropriate approaches, e.g. our IISAN-Versa, can effectively and efficiently improve multimodal recommendation.
    \item We construct a new multimodal text recommendation scenario by extending the open-source MicroLens dataset, i.e., generating captions from its original videos and images using a pre-trained video captioning generator, to validate the multi-scenario adaptation capabilities of the proposed IISAN-Versa. The new reconstructed dataset will be released to facilitate future research for the multimodal recommendation community.
\end{itemize}

\section{Related Work}
\textbf{Multimodal Foundation Models}. Recent work in multimodal learning focuses on leveraging pre-trained models for downstream tasks to enhance performance and reduce pre-training costs \cite{chen2024knowledge}. BERT \cite{devlin2018bert} pioneered the pretraining and fine-tuning paradigm in NLP, using Masked Language Modeling and the Next Sentence Prediction to understand context and sentence relationships. Following this, the Vision Transformer (ViT) \cite{dosovitskiy2020image} adapted transformer architecture for image classification, demonstrating competitive performance by treating image patches as tokens. CLIP~\cite{radford2021learning} further bridged vision and language, training a multimodal foundation model to match images with their textual descriptions through contrastive learning, enabling powerful zero-shot learning capabilities \cite{radford2021learning}. The evolution continued with large-scale models like GPT-4, and multimodal variants such as Flamingo \cite{alayrac2022flamingo}. These models, characterized by their vast scale and diverse training data, perform a wide array of tasks, pushing the boundaries of AI's multimodal understanding and processing capabilities. 

\noindent \textbf{Modality-based Sequential Recommendation (MoRec)}.
The Recommender System community has shown a growing interest in incorporating various modality information into recommendation systems \cite{fu20251st,yuan2023go,cheng2023image,ni2023content,fu2025crossan}. These systems utilize large-scale multimodal foundation models \cite{devlin2018bert,radford2021learning,he2025double} from NLP and CV \cite{dosovitskiy2020image,liu2021swin} to encode text and image data. The sequential encoder remains consistent with traditional architectures like SASRec \cite{kang2018self}, GRU4Rec \cite{hidasi2015session}, and BERT4Rec \cite{sun2019bert4rec}. Furthermore, models such as IDA-SR \cite{mu2022id}, UniSRec \cite{hou2022towards}, and VQ-Rec \cite{hou2022learning} have advanced MoRec by developing item representations using NLP foundation models. For image representation in recommender systems, studies such as \cite{wei2019mmgcn} and \cite{meng2020heterogeneous} have integrated image features extracted from ResNet-50 \cite{He2016deep} into the models.

Furthermore, research \cite{yuan2023go,ni2023content,li2023multi} has shown that the MoRec framework, applied through end-to-end learning, significantly outperforms previous methods that rely on offline feature extraction. Specifically, \cite{li2023multi, li2024empirical} emphasized that end-to-end training, which integrates both image and text modalities, significantly surpasses the methods that rely solely on a single modality. Although directly learning the raw content showcases tremendous performance, a significant limitation of these end-to-end studies is their ongoing reliance on fully fine-tuning large-scale multimodal encoders, which often leads to inefficiency.

\noindent \textbf{Parameter-efficient Fine-tuning (PEFT)}. In the fields of Natural Language Processing (NLP) and Computer Vision (CV), substantial research efforts are focused on investigating Parameter-Efficient Fine-Tuning (PEFT) methods. These methods aim to tackle the challenge posed by the vast number of trainable parameters in large-scale pre-trained models. Foundational works such as Adapter~\cite{houlsby2019parameter}, LoRA~\cite{hu2021lora}, and BitFit~\cite{zaken2021bitfit} have laid the groundwork in this area. Building on these foundational works, a variety of alternative strategies have emerged, as seen in studies like \cite{wang2020k}. Despite the advancements, these approaches primarily utilize Embedded PEFT (EPEFT) and focus largely on parameter efficiency, often overlooking practical efficiency concerns such as training speed and GPU memory usage.

To overcome these limitations, \cite{cai2020tinytl} proposed the concept of \textcolor{black}{``}Reduce Memory, Not Parameters," advocating for a reduction in activations. Additionally, LST \cite{sung2022lst} offers a memory-efficient strategy aimed at decoupling PEFT modules. A notable shift in recent NLP and CV research is the transition from an emphasis on Embedded PEFT (EPEFT) to Decoupled PEFT (DPEFT), as illustrated by recent works \cite{xu2023side,lin2023vision,xu2023san}. This shift reflects an evolving understanding that focuses more on practical efficiency instead of trainable parameters.

In modality-based sequential recommender systems (RS), PEFT methods have also shown significant progress, as demonstrated by M6-Rec~\cite{cui2022m6}, and Tallrec~\cite{bao2023tallrec}. These studies indicate that PEFT approaches can achieve performance levels comparable to traditional fine-tuning methods. However, most such methods still rely on conventional EPEFT, often ignoring practical efficiency aspects. The adoption of DPEFT in recommender systems remains limited. Moreover, these methods generally focus on single-modality analysis, missing opportunities to leverage the abundant multimodal information in recommendation scenarios. VIP5~\cite{geng2023vip5} investigates multimodal recommendations using adapters within the P5 framework, though it primarily utilizes adapters for text encoders and leaves image encoders unchanged for feature extraction. This contrasts with our approach, as adapting multiple modality encoders incurs greater costs than only using the text encoder. We initially introduced IISAN \cite{fu2024iisan}, an efficient Decoupled PEFT approach designed for symmetrical multimodal foundational models in sequential recommendation. However, it has limitations in effectively handling asymmetrical MFMs, preventing it from leveraging the capabilities of recent state-of-the-art large language models.

Overall, little research has delved into the practical efficiency challenges of using DPEFT approaches for versatile multimodal representation adaptation in recommendation tasks. While multimodal information can significantly enhance recommendation performance \cite{li2023multi}, it also intensifies practical efficiency challenges. Consequently, investigating DPEFT strategies to address real efficiency issues in this domain represents a novel and pressing research direction. Although IISAN partially resolves this problem, it still suffers from the limitation of incorporating asymmetrical MFMs.  In this paper, we extend IISAN into a more versatile framework that improves its ability to incorporate large and asymmetrical MFMs.

\section{Methodology}

\subsection{Overview}
\begin{figure}
  \centering
\includegraphics[width=\linewidth]{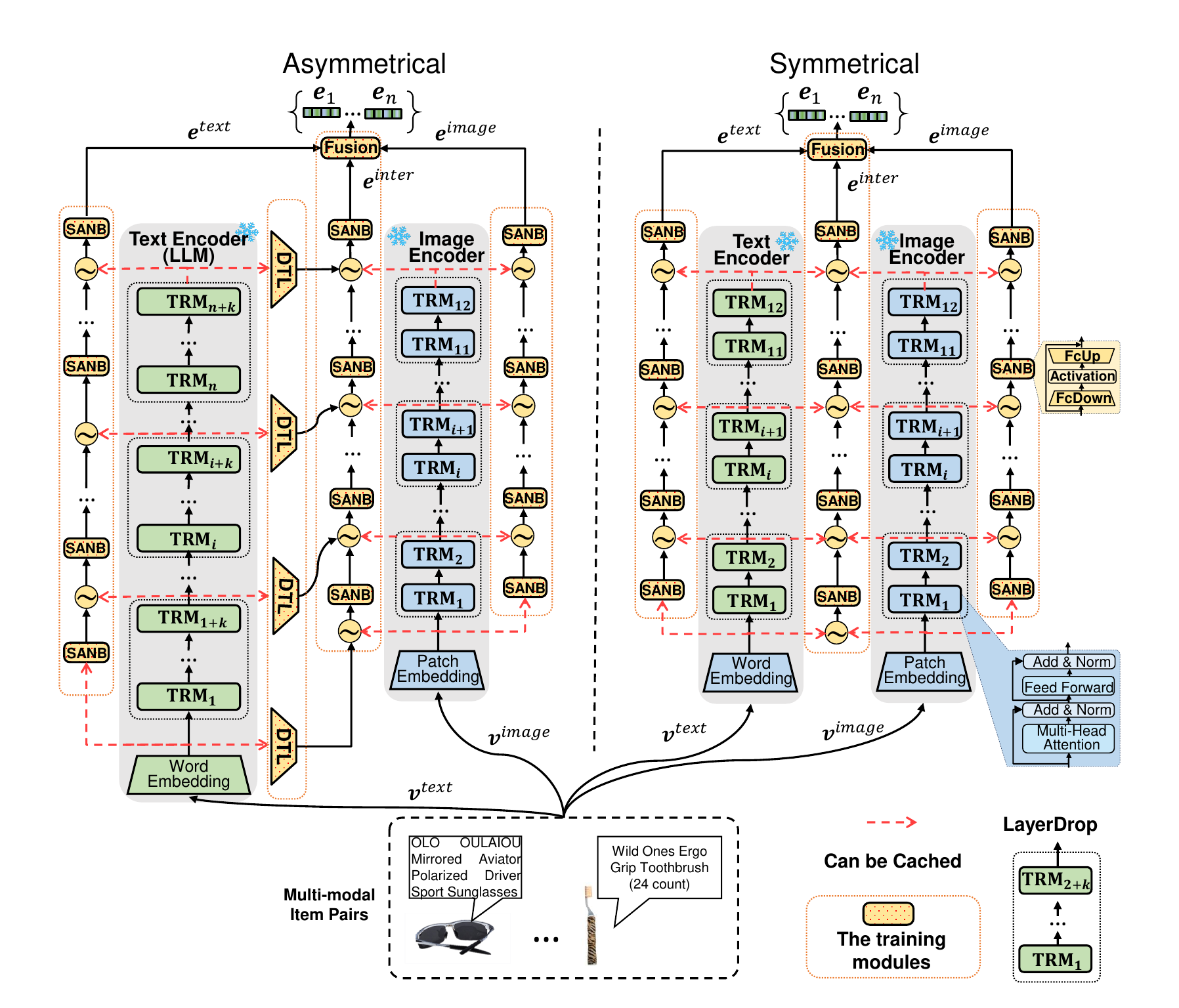}
\vspace{-0.1in}
  \caption{An Overview of the IISAN-Versa for Sequential Recommendation. The IISAN-Versa model introduces versatile intra- and inter-modal side-adapted networks (SANs) designed for adapting symmetrical and asymmetrical multimodal foundation model in sequential recommendation. These networks consist of multiple SAN blocks (SANBs) and learnable fusion gates, merging two asymmetrical structures. IISAN-Versa employs dimension transformation layers (DTLs) to align the embedding dimensions. Notably, we leverage LayerDrop to further eliminate redundancy. The asymmetrical framework utilizes an even-numbered group LayerDrop technique, which drops K blocks per group. This approach ensures that the large text encoder aligns seamlessly with the image encoder. }
    \label{fig:iisan_versa} 
    \vspace{-0.2in}
\end{figure}

The proposed IISAN-Versa framework, shown in Figure \ref{fig:iisan_versa}, is designed to adapt both symmetrical and asymmetrical structures of multimodal foundation models, enabling versatile representation generation. By separating the new trainable side-adapted networks (SAN) from multimodal backbones, we optimize the backpropagation computation graph, addressing practical efficiency in adapting large-scale multimodal foundation models to downstream tasks, which will be further explained in Section \ref{sec:eff_analysis}. Leveraging the decoupling features of IISAN-Versa, the hidden states of the multimodal foundation models can be cached to enhance IISAN-Versa's efficiency. The architecture introduces independent intra-modal SANs for visual and textual modalities and an inter-modal SAN for managing their interactions. This framework allows the efficient adaptation of pre-trained MFMs like BERT, Llama, and ViT to downstream multimodal recommendation tasks.

In this paper, IISAN-Versa's innovation lies in proposing versatile intra- and inter-modal side-adapted networks to efficiently and effectively adapt both symmetrical and asymmetrical multimodal foundation models. For symmetrical MFMs, IISAN-Versa adopts balanced adaptation following IISAN~\cite{fu2024iisan}. The main challenge tackled in this paper is the adaptation of asymmetrical structures of MFMs, which present two primary issues: (1) inconsistent embedding dimensions and (2) imbalanced hidden layers. This paper is dedicated to proposing simple yet effective solutions to resolve them. To address the first issue, we adopt a dimension transformation layer to align the differences between the large language model (LLM) and the smaller image encoder. For the second issue, we employ a group LayerDrop strategy to resolve the imbalance in hidden layers.

\subsection{Problem Formulation} \label{sec:problem}
Given a recommendation dataset $\mathcal{D} = \{\mathcal{U}, \mathcal{V}\}$, where $\mathcal{U}$ and $\mathcal{V}$ denote the set of users and the set of items, respectively, we aim to predict the next item interacted with by a user $u$ in a multimodal sequential recommendation task by exploiting their $n$ past behaviors. In multimodal recommendation, each item $v$ contains two modal representations, i.e., text ($v^{\text{t}}$) and corresponding image ($v^{\text{v}}$). These texts and images are fed into two pre-trained multimodal foundation models, such as BERT \cite{devlin2018bert} or Llama \cite{touvron2023llama} for texts and ViT \cite{dosovitskiy2020image} for images, which consist of an embedding layer (word embedding for texts and patch embedding for images) and multiple transformer blocks, respectively. Through these pre-trained multimodal backbones, we obtain textual and visual embeddings from the embedding layers and multiple hidden states ($\{h_i^{\text{t}}\}$ and $\{h_i^{\text{v}}\}$) from the transformer blocks ($\{TRM_i\}$), where $i$ denotes the $i$-th layer of the backbone. Specifically, for IISAN-Versa, we decouple the trainable parameters into three separate towers: the textual-modality side-adapted network for text representation, obtaining the final text embedding \textcolor{black}{$e^{\text{t}}$}; the visual-modality side-adapted network for image representation, obtaining the image embedding \textcolor{black}{$e^{\text{v}}$}; and an inter-modal side-adapted network for image-text interactive representation, obtaining the inter-modality embedding $e^{\text{inter}}$. 
A linear-based fusion layer $(FL)$ is added to ensure the consistency of the output dimensions of the item embedding and input dimensions of the sequential encoder following \cite{hou2022towards, yuan2023go} for the final recommendation, as follows: 
\textcolor{black}{
\begin{equation}
{e^{item}} = FL([{e^{v}} : {e^{inter}}:{e^{t}}])
\end{equation}
}
where $[:]$  means the feature concatenation. Then, we input the $\boldsymbol{e^{item}}$ into the sequential encoders and calculate the final predicted score for user $u$ to $i$-th item as $\boldsymbol{\hat{y}_{ui}}$, which is the product of the output sequential encoder and corresponding item embedding. 

\vspace{-0.1in}
\subsection{Intra-model Side Adapted Network}
Each side-adapted network (SAN) consists of several SAN blocks (SANBs), which include Upsample and Downsample layers built on a fully connected network. Each SANB, a unit within the SAN, serves as a network block. In this research, we adopted the adapter block as the basic unit of the SAN, termed SANB. Unlike the insertion from Adapter Tuning~\cite{houlsby2019parameter}, SANB is implemented in a decoupled manner. For instance, in the textual SAN, a learnable gate mechanism is utilized to fuse the information \(h_{i-1}^{B^{intra}}\) from the previous SANB with the current hidden state \(h_i^{t}\), as demonstrated in the following formula:
\textcolor{black}{
\begin{equation}
    h_i^{B^{intra}} = SANB^{intra}\left(\mu_i^{t} \ast h_{i-1}^{B^{intra}} + (1 - \mu_i^{t}) \ast h_i^{t}\right)
\end{equation}
} 
where $\mu_i^{t}\in [0,1]$. Note that the first SANB will only take the text embeddings and the visual SAN employs similar operations. This Intra-SAN is applicable to both symmetrical and asymmetrical frameworks, the only difference we should make is to enlarge the dimension of input and output dimensions of SANB to fit the larger encoders.

\subsection{Inter-model Side Adapted Network}
When adapting IISAN-Versa to asymmetrical multimodal foundation models, the key difference is in handling the fusion of inter-modality. By performing group layer dropping, we transform the asymmetrical number of hidden states into a symmetrical form, as explained in Section \ref{sec:layerdrop}. \textcolor{black}{Following prior work~\cite{fu2024exploring,yuan2023go}, we adopt a \textit{Dimension Transformation Layer (DTL)} to align the high-dimensional embeddings from the LLM with the lower-dimensional outputs of vision encoders. The DTL is computed as:
\[
\mathbf{z} = \mathbf{W} \mathbf{x} + \mathbf{b}, \quad \mathbf{W} \in \mathbf{R}^{d_{\text{out}} \times d_{\text{in}}},\ \mathbf{b} \in \mathbf{R}^{d_{\text{out}}}
\]
where \( \mathbf{x} \in \mathbf{R}^{d_{\text{in}}} \) is the input feature, \( \mathbf{z} \in \mathbf{R}^{d_{\text{out}}} \) is the transformed output, \( d_{\text{in}} \) and \( d_{\text{out}} \) denote the input and output dimensions respectively, and \( d_{\text{in}} > d_{\text{out}} \).
}

We then apply a fusion mechanism to merge the hidden states of the two modalities and sum the information from the last SANB using a gating mechanism similar to Intra-SAN, as described by the following formula:
\begin{equation}
    h_i^{B^{inter}} = SANB^{inter}\left(\beta_i h_i^{v} + (1 - \beta_i) X_i + h_{i-1}^{B^{inter}}\right)
\end{equation}
where
\begin{equation}
X_i =
\begin{cases}
h_i^{t}, & \text{for IISAN-VS} \\
DTL(h_i^{t}), & \text{for IISAN-VA}
\end{cases}
\end{equation}

\begin{algorithm}[h]
\caption{\textcolor{black}{Group LayerDrop for layer selection}}
\label{alg:select_layers}
\begin{algorithmic}[1]
\Function{SelectLayers}{$L, L^{v}$}
    \State $step \gets \lceil L / L^{v} \rceil$
    \State $S \gets [\,]$
    \For{$i = L^{v}-1$ \textbf{to} $0$}
        \State \textbf{append} $L - step \times i$ \textbf{to} $S$
    \EndFor
    \State \Return $S$

\EndFunction
\end{algorithmic}

\end{algorithm}

\begin{table}[htbp]
\centering
\caption{\textcolor{black}{Selected layers \(S\) for various total depths \(L\) (with \(L^{v}=6\)).}}
\label{tab:selected_layers}
\renewcommand\tabcolsep{14pt}
\scalebox{0.9}{
\begin{tabular}{cc}
\hline
\textcolor{black}{\multirow{2}{*}{Number of Layers}}&\textcolor{black}{\multirow{2}{*}{Selected layers}}  \\
\\
\hline
\textcolor{black}{
12} & \textcolor{black}{\(\{2,4,6,8,10,12\}\)} \\
\textcolor{black}{24} & \textcolor{black}{\(\{4,8,12,16,20,24\}\)} \\
\textcolor{black}{28} & \textcolor{black}{\(\{3,8,13,18,23,28\}\)} \\
\textcolor{black}{32} & \textcolor{black}{\(\{2,8,14,20,26,32\}\)} \\
\textcolor{black}{48} & \textcolor{black}{\(\{3,12,21,30,39,48\}\)} \\
\textcolor{black}{80} & \textcolor{black}{\(\{5,20,35,50,65,80\}\)} \\

\hline
\end{tabular}
}
\vspace{-0.1in}
\end{table}

where $\beta_i\in [0,1]$.  Note that the first inter-SANB only inputs the text embedding and the visual embeddings. We define the symmetrical and asymmetrical IISAN-Versa as IISAN-VS and IISAN-VA.

\subsection{LayerDrop} \label{sec:layerdrop}
Additionally, to further enhance network efficiency and address issues of layer redundancy, we introduce LayerDrop techniques, similar to those in \cite{sung2022lst,fu2024iisan}, to reduce the number of SAN blocks and align asymmetrical MFM. Specifically, for symmetrical MFM, we group two transformer blocks together and drop the first hidden state to the SANs, which can save half the number of SANBs. In the case of asymmetrical MFM, to address the imbalance between the layers of LLM and vision encoders, we apply a simple yet effective group LayerDrop technique that groups $k$ layers evenly from the hidden states of the LLM. Suppose there are $L^{\text{t}}$ and $L^{\text{v}}$ layers in the text encoder and image encoder, respectively. In scenarios of asymmetrical adaptation, the text encoder is an LLM, thus $L^{\text{t}} \gg L^{\text{v}}$. For image encoders, we continue using the same LayerDrop technique as with symmetrical structures, where we drop half the number of layers. In this paper, we design to select the largest $k$ to ensure the maximum range of cross-layer interaction within the entire LLM. The number $k$ can be represented as follows:
\begin{equation}
    \max_{k} \left( L^{\text{t}} - k \cdot \frac{L^{\text{v}}}{2} \geq 1 \right)
\end{equation}
\textcolor{black}{
We compute
\(
\mathrm{step} = \bigl\lceil \tfrac{L}{L^{v}} \bigr\rceil
\)
and select layers as shown in Algorithm~\ref{alg:select_layers}.
The specific selections for various \(L\) (with \(L^{v}=6\)) are presented in Table~\ref{tab:selected_layers}.} In Section \ref{sec:ablation}, we further investigate various LayerDrop schemes, showcasing the performance impact of dropping different numbers of blocks in both symmetrical and asymmetrical structures.

\subsection{Caching Strategies}
We adopt a caching technique as a refinement strategy for efficiency~\cite{karedla1994caching,chen2020mobile}. As shown in Figure \ref{fig:iisan_versa}, the decoupled PEFT mechanism, consisting of the separable pre-trained backbone model and the new trainable model, enables this approach. This technique involves the storage and reuse of item hidden states extracted from pre-trained multimodal backbones, thereby minimizing the need for repeated forward passes of foundational models during training. This method can significantly enhance efficiency for asymmetrical IISAN-Versa. The primary reason is that it fundamentally avoids the necessity of moving the LLM into the GPUs, thus making it feasible to involve the LLM in the adaptation process even with very limited resources. Note that all EPEFT methods are not able to be cached due to the nature of embedding trainable parameters inside the foundation models, which will be further explained in Section \ref{sec:eff_analysis}.

\subsection{Training Objectives}
In terms of training details, we exploit the in-batch debiased Cross-Entropy loss function $\mathcal{L}_{CE}$~\cite{yuan2023go,ni2023content} widely adopted in both academic literature and industrial systems.
\begin{equation}
D_{ui} = \exp(\hat{y}_{ui} - \log(p_i)) + \sum_{j \in [B], j \notin I_u} \exp(\hat{y}_{uj} - \log(p_j))
\end{equation}
\begin{equation}
\mathcal{L}_{CE}=-\sum_{u \in \mathcal{U}} \sum_{i \in [2,...,n+1]} \log \frac{\exp(\hat{y}_{ui} - \log(p_i))}{D_{ui}}
\end{equation}
where  $p$ is the popularity of the item, $I_u$ and $B$ stand for the item set interacted by user $u$ and the batch. $n+1$ item denotes the predicted item for user $u$.  $D_{ui}$ is a temporary variable to simplify Formula (5).

\section{Efficiency Analysis}\label{sec:eff_analysis}

During the training of a neural network, the computation involved in backward propagation and the forward pass often represents a significant bottleneck in terms of both training time and GPU memory usage. In this section, we provide an efficiency analysis to understand the differences in the forward and backward propagation processes of full fine-tuning (FFT), Embedded Parameter-Efficient Fine-Tuning (EPEFT) methods such as Adapter and LoRA, and the Decoupled PEFT, IISAN-Versa, in both its Uncached and Cached approach. We argue that traditional embedded PEFT does not substantially improve training efficiency because it does not significantly reduce the computational load of backward propagation. In contrast, the Uncached IISAN-Versa approach can save computation in the backward pass, while the Cached IISAN-Versa can also save time during the forward pass of large foundation models.

To simplify the theoretical analysis, we consider a scenario involving large foundation models comprising $n$ transformer blocks, where the $i$-th transformer block is denoted by $TRM_i$. For each transformer block, all PEFT methods, including Adapter, LoRA, and IISAN, will place a trainable PEFT module denoted by $peft$. We assume that $TRM \gg peft$ according to the definition of parameter efficiency.

In the forward pass of full fine-tuning, we denote the final output as $Z_n$. This process can be represented as follows:
\begin{equation}\label{eq:fp_trm}
    Z_n = TRM_n(\cdots TRM_2(TRM_1(x)))
\end{equation}
To update each neural network module, it is necessary to calculate the gradients, starting from the initial block $TRM_1$. This leads to the longest propagation chain of the loss $L$ with respect to the parameters of the first block:
\begin{equation}
\frac{\partial L}{\partial TRM_1} = \frac{\partial L}{\partial Z_n} \cdot \frac{\partial Z_n}{\partial Z_{n-1}} \cdot \frac{\partial Z_{n-1}}{\partial Z_{n-2}} \cdots \cdot \frac{\partial Z_2}{\partial Z_1} \cdot \frac{\partial Z_1}{\partial TRM_1}
\end{equation}

For embedded PEFT, the modules are incorporated inside each transformer block. This can be expressed as:
\begin{equation}
    p_n = peft_n(TRM_n(\cdots peft_2(TRM_2(peft_1(TRM_1(x)))))
\end{equation}
We denote the output of the $n$-th PEFT module as $p_i$ and the output of the $n$-th transformer block as $Z^*_n$:
\begin{equation}
    Z^{*}_n = TRM_n(\cdots peft_2(TRM_2(peft_1(TRM_1(x))))
\end{equation}
Thus, the overall forward and backward propagation through the entire foundation model is expressed as:
\begin{equation}
\frac{\partial L}{\partial peft_1} = \frac{\partial L}{\partial p_n} \cdot \frac{\partial p_n}{\partial Z^{*}_{n}} \cdot \frac{\partial Z^{*}_{n}}{\partial p_{n-1}} \cdot \cdots \cdot \frac{\partial Z^{*}_2}{\partial p_1} \cdot \frac{\partial p_1}{\partial peft_1}
\end{equation}
The additional computational cost introduced by PEFT modules is generally negligible because $peft \ll TRM$. However, in practice, embedded PEFT can still offer efficiency improvements by reducing the gradients stored on the GPU and minimizing training time during weight updates, as demonstrated in \cite{fu2024iisan}. Nonetheless, the primary computational challenges associated with both forward and backward propagation in large foundation models remain.

In contrast, for decoupled PEFT approaches like IISAN-Versa, the training process differs significantly. The decoupled framework splits the forward pass into two stages. The first stage involves a forward pass through the backbone model, as shown in Equation \ref{eq:fp_trm}. However, the key difference is that the outputs (hidden states) of each transformer block layer are cached, allowing these outputs to serve as constant inputs to the standalone PEFT modules. The forward pass of decoupled PEFT modules is then defined as:
\begin{equation}\label{eq:fp_dpeft}
    p^{*}_n = peft_n(\cdots peft_2(peft_1(x)))
\end{equation}
Thus, the backward pass does not need to traverse the large foundation model:
\begin{equation}\label{eq:bp_dpeft}
\frac{\partial L}{\partial peft_1} = \frac{\partial L}{\partial p^{*}_n} \cdot \frac{\partial p^{*}_n}{\partial p^{*}_{n-1}} \cdot \frac{\partial p^{*}_{n-1}}{\partial p^{*}_{n-2}} \cdots \cdot \frac{\partial p^{*}_2}{\partial p^{*}_1} \cdot \frac{\partial p^{*}_1}{\partial peft_1}
\end{equation}

For IISAN-Versa (Cached), hidden states are stored in on-disk memory rather than recalculated each time during training, further optimizing the forward pass and achieving a high level of efficiency. Therefore, we can further give the definition of DPEFT, which is the PEFT method whose forward pass can be written as the form of Equation \ref{eq:fp_trm} and Equation \ref{eq:fp_dpeft} with the backward pass only by Equation \ref{eq:bp_dpeft}, where the step of Equation \ref{eq:fp_trm} can be optimized by caching. 

In this analysis, we compared Full Fine-Tuning (FFT), Embedded PEFT, and Decoupled DPEFT in terms of their efficiency and resource utilization when training large foundation models. We found that EPEFT does not significantly boost efficiency because it does not effectively reduce computational demands in both forward and backward propagation. This approach still requires traversing the entire model during training, which limits its potential to save computational resources. In contrast, DPEFT (IISAN-Versa) reduces computational demands by decoupling the fine-tuning process, significantly minimizing backward propagation through the foundation models. Moreover, when equipped with a caching strategy, DPEFT can handle forward propagation more efficiently by performing it once and caching the results for subsequent use. This combination makes DPEFT a highly efficient and resource-effective approach for adapting large foundation models.

\section{Experiment Setup}
In this section, we illustrate the used datasets, caption generation, evaluation methods, and implementation details of the IISAN-Versa.

\begin{table}
  \caption{Dataset Description}
  \renewcommand\tabcolsep{5.3pt}
  \renewcommand{\arraystretch}{0.9}
  \label{tab:dataset}
  \scalebox{0.9}{
  \begin{tabular}{clcccc}
    \hline
    \multirow{2}{*}{Dataset}&\multirow{2}{*}{Users}&\multirow{2}{*}{Items}&\multirow{2}{*}{Interaction}&\multirow{2}{*}{Content}\\
    &&&&&\\
    \hline
    Scientific&12,076&20,314&81,711&Text\&Image\\
    Instrument&10,000&19,246&75,022&Text\&Image\\
    Office&10,000&22,785&71,276&Text\&Image\\
    MicroLens&100,000&19,738&719,405&Text\&Image\&video\\
    \textcolor{black}{NINERec-1M}&\textcolor{black}{151,495}&\textcolor{black}{104,605}&\textcolor{black}{1,035,822}&\textcolor{black}{Text\&Image}\\
  \hline
\end{tabular}
}
\vspace{-0.15in}
\end{table}

\subsection{Dataset}\label{sec:dataset}
In this study, each item in a user sequence is represented by its raw modality information. To assess our methods on items that include both image and text modalities, we utilize the Amazon review datasets \cite{ni2019justifying}. Specifically, we employ three widely used datasets from this collection: \textcolor{black}{``}Industrial and Scientific," \textcolor{black}{``}Musical Instruments," and \textcolor{black}{``}Office Products." These datasets were adopted following the preprocessing steps outlined in prior works \cite{fu2024iisan,yuan2023go,fu2024exploring,li2023exploring}. Furthermore, to explore the performance of IISAN-Versa compared with other recommendation approaches with a fair comparison, we also incorporate MicroLens \cite{ni2023content}, a public multimodal video recommendation benchmark that includes raw text, images, and video content. Since MicroLens is a dataset rich in modality information, we also explore multimodal text scenarios within this public benchmark. \textcolor{black}{To improve the evaluation scale, we additionally sampled one million interactions from the publicly available NINERec dataset \cite{zhang2023ninerec}.}

\begin{figure}
  \centering \includegraphics[width=0.9\linewidth]{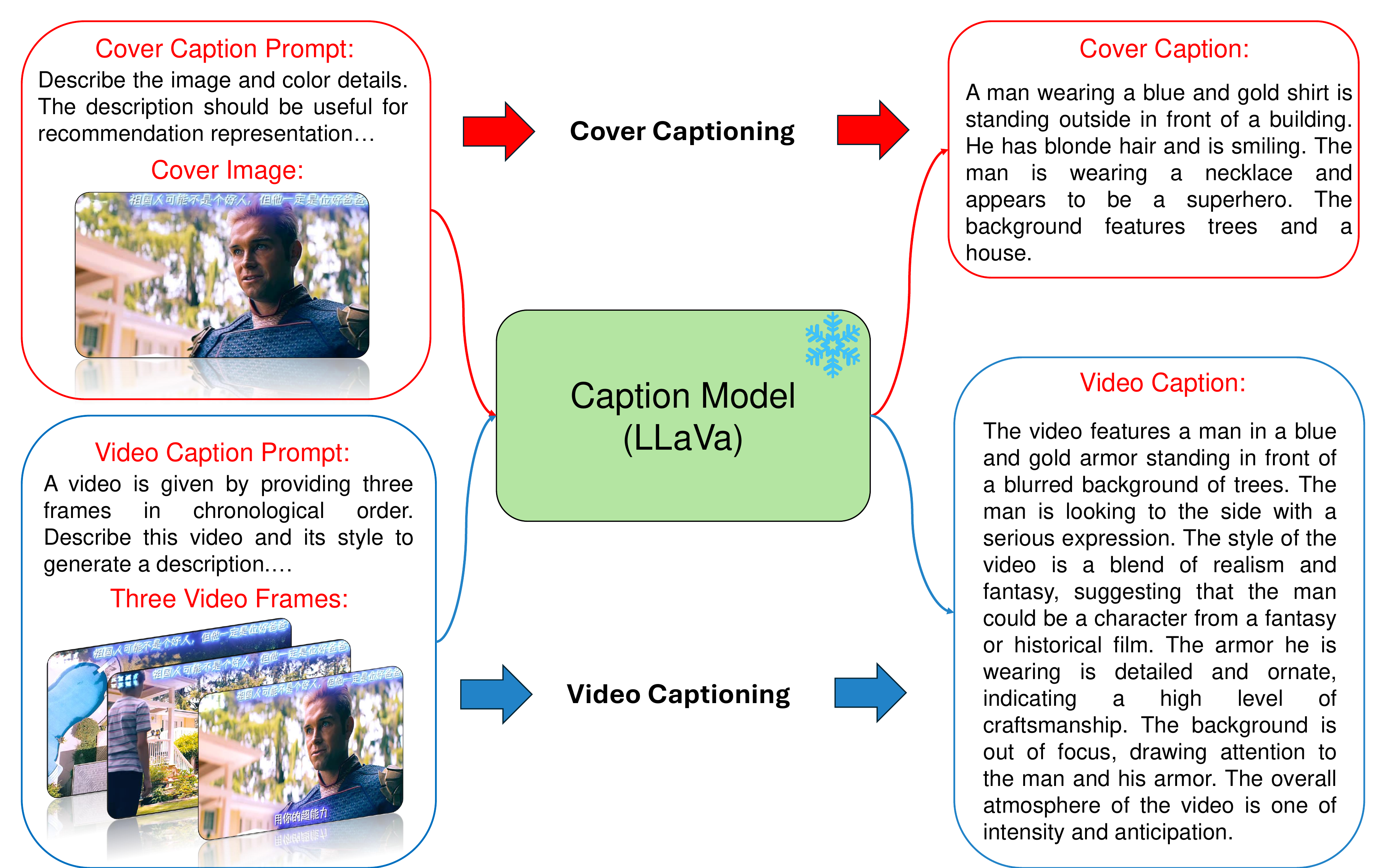}
  \caption{An Overview of Caption Generation. The image or video frames, along with the corresponding caption prompt, are input into the pre-trained LLaVA caption model. The model, kept frozen during this process, generates both the cover and the caption simultaneously.}
    \label{fig:caption} 
    \vspace{-0.15in}
\end{figure} 

\subsection{Multimodal Text Generation} \label{sec:caption}
To explore the potential of IISAN-Versa in adapting multiple scenarios, such as multimodal text scenarios containing cover text and corresponding video texts, we utilize the pretrained LLaVA model to generate captions for image and video content from the public benchmark MicroLens datasets, as illustrated in Figure \ref{fig:caption}. For this process, we input the cover image and extract three evenly spaced frames from each video. In terms of caption models, we use \textcolor{black}{``}llava-hf/llava-v1.6-mistral-7b-hf" for cover image captioning. For video captioning, we apply \textcolor{black}{``}liuhaotian/llava-v1.6-34b" due to the higher complexity of videos; a larger caption model provides a more accurate representation of the video content. Our video captioning strategies follow the preprocessing methods used in Video generation OpenSora\footnote{\url{https://github.com/hpcaitech/Open-Sora}}. To enhance LLaVA's ability to produce effective representations for postprocessing, we incorporate in-context learning by including concrete examples in the prompt. We adopt the prompt from the Sora project\footnote{\url{https://openai.com/sora}} for video prompts and use the prompt from DALLE2\footnote{\url{https://openai.com/index/dall-e-2}} for cover image prompts.

Finally, these multiple generated captions as well as the original texts will be used as the multimodal text input to IISAN-Versa to evaluate multi-scenario generalization ability. 

\vspace{-0.1in}
\subsection{Performance Evaluations.}
Based on previous research \cite{he2017neural, yuan2023go, fu2024exploring, fu2024iisan}, our methodology employs a leave-one-out evaluation strategy. In this approach, the final item in the interaction sequence is reserved for testing, the penultimate item is used for validation, and all preceding items are utilized for training. We evaluate our model using HR@10 (Hit Ratio) and NDCG@10 (Normalised Discounted Cumulative Gain), consistent with the evaluation methodologies in \cite{he2017neural, yuan2023go, fu2024exploring}, as the key performance metrics. Unless stated otherwise, all results reported refer to the test set. Furthermore, it is crucial to mention that the predicted item is evaluated against the complete set of items.
\subsection{Implementation Details.} 
 The pretrained models from the Huggingface platform\footnote{https://huggingface.co/}, detailed in Table \ref{tab:model_specs}, are employed as text and image encoders in this research. Note that due to the limited computing resources, we adopt the Llama-3-70b with GPTQ 4-bit quantization. We experiment with hidden representation dimensions for the sequential encoder from \{32, 64, 128\} and set it to 64. We fixed the sequential encoder to include 2 Transformer blocks and 2 attention heads following \cite{fu2024iisan,yuan2023go}. 
 \begin{table}[htbp]
\centering
\caption{Description of the pre-trained models used in this research. L denotes the number of Transformer blocks, H represents the hidden dimension of the transformer layer, B indicates the number of parameters in billions, and M indicates the number of parameters in millions.}
\label{tab:model_specs}
\renewcommand{\arraystretch}{0.9}
\renewcommand\tabcolsep{3pt}
\scalebox{0.8}{
\begin{tabular}{cccc}
\hline
\multirow{2}{*}{Modality} & \multirow{2}{*}{Pre-trained model} & \multirow{2}{*}{Architecture} & \multirow{2}{*}{\#Param} \\
\\
\hline
\multirow{13}{*}{Text} 
 & BERT-tiny & L=2, H=128 & 4.4M \\
 & BERT-base & L=12, H=768 & 110M \\
 & BERT-large & L=24, H=1024 & 340M \\
 & Gemma-7b & L=28, H=3072 & 7B \\
 & Gemma-7b-it & L=28, H=3072 & 7B \\
 & Mistral-7b & L=32, H=4096 & 7B \\
 & Llama2-7b & L=32, H=4096 & 7B \\
 & Llama2-7b-chat & L=32, H=4096 & 7B \\
 & Llama3-8b & L=32, H=4096 & 8B \\
 & Llama3-8b-it & L=32, H=4096 & 8B \\
 & Mixtral-8x7b-it & L=32, H=4096 & 47B \\
 & Mixtral-8x7b & L=32, H=4096 & 47B \\
 & Llama3-70b-GPTQ-it & L=80, H=8192 & 70B \\
 & Llama3-70b-GPTQ & L=80, H=8192 & 70B \\
\hline
\multirow{3}{*}{Image} 
 & ViT-tiny & L=12, H=192 & 21M \\
 & ViT-base & L=12, H=768 & 86M \\
 & ViT-large & L=24, H=1024 & 307M \\
 & ViT-huge & L=32, H=1280 & 632M \\
  & EVA-ViT & L=48, H=5120 & 11B \\
\hline
\end{tabular}}
\vspace{-0.1in}
\end{table}
 Using the Adam optimizer without weight decay, we tested learning rates from 1e-5 to 1e-3 while maintaining a dropout probability of 0.1. We explored batch sizes from \{8, 16, 32, 64, 512\}, selecting the largest viable option to optimize GPU memory, and searched for SANB hidden dimensions from \{8, 16, 32, 64, 128, 512, 1024, 2048\}. Hyperparameters were searched separately for the MicroLens and Amazon datasets. All hyperparameters were finalized based on validation performance, and results were reported using the test set. The experiments were conducted on one A6000 GPU.

\section{Experiment}
In this section, we explore IISAN-Versa by addressing the following research questions:
\begin{itemize}
    \item \textbf{RQ1:} Does Asymmetrical IISAN-Versa's performance improve when equipped with a larger text encoder compared to its standard symmetrical IISAN-Versa?
    \item \textbf{RQ2:} If the answer to RQ1 is ''Yes'', does IISAN-Versa exhibit a scaling effect with the integration of progressively larger LLMs?
    \item \textbf{RQ3:} How does IISAN-Versa compare in performance against established public benchmarks?
    \item \textbf{RQ4:} How is the efficiency of IISAN-Versa compared with popular EPEFT approaches?
    \item \textbf{RQ5:} What are the key findings about the components affecting the performance of IISAN-Versa?
    \item \textbf{RQ6:} Would IISAN-Versa show advantages in processing other multimodal scenarios such as multimodal text using both raw text (title) and the text generated (cover and video captions) from images or video?
    \item \textcolor{black}{\textbf{RQ7:} How does the IISAN-Versa perform with different hyper-parameters?}
\end{itemize}

\vspace{-0.1in}
\subsection{IISAN-Versa Performance with larger text encoders (RQ1)}
\label{sec:main}

\begin{figure}
  \centering
   \includegraphics[width=0.9\linewidth]{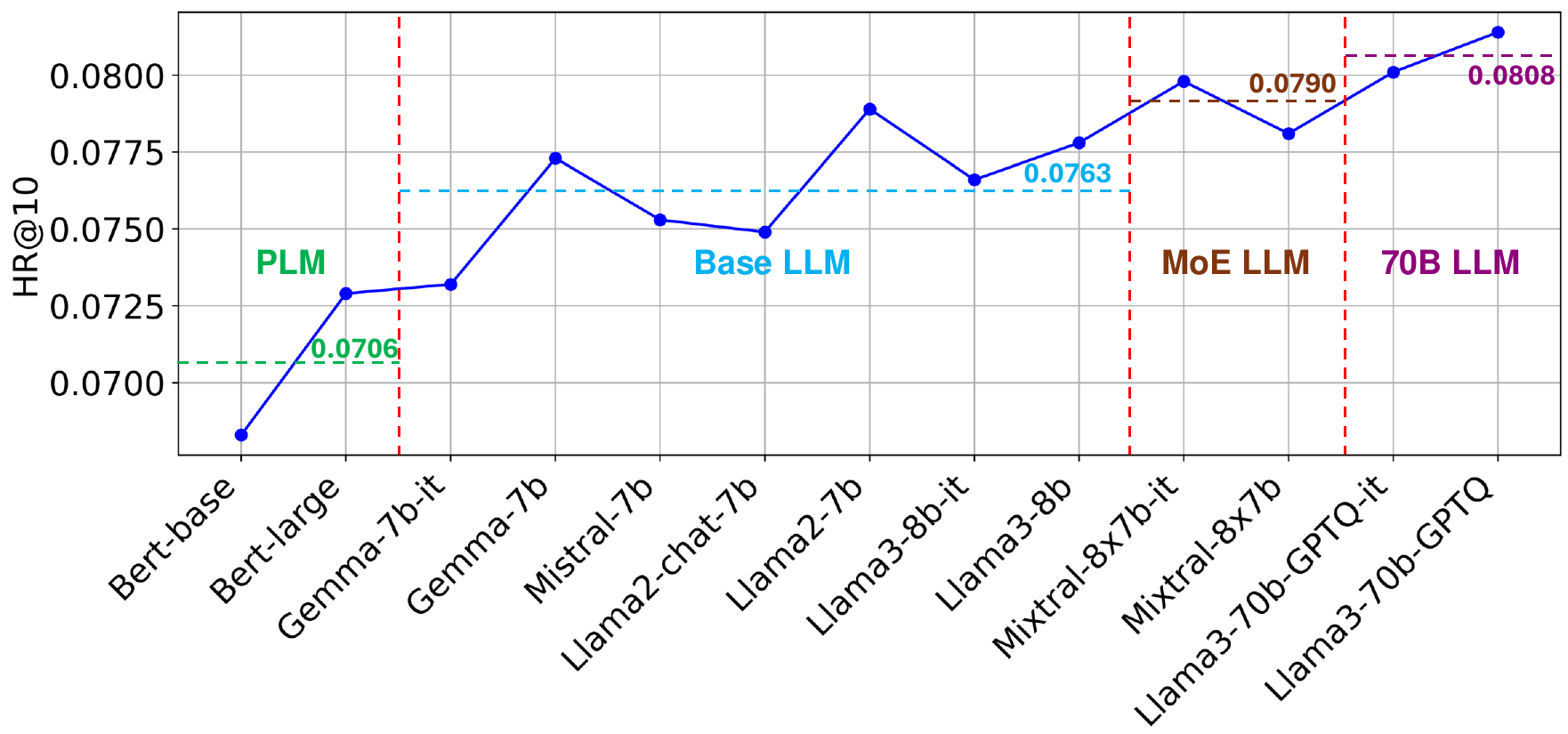}
   \vspace{-0.1in}
  \caption{Scaling effect of IISAN-Versa's text encoder on Scientific dataset with Hit@10 metric. We group the pre-trained text encoders based on their scale, with each group increasing in size from left to right. We adopt the vertical red dotted line to separate each group. The horizontal line represents the average performance of each group.}
    \label{fig:scaling} 
    \vspace{-0.1in}
\end{figure}

\begin{table}[htbp]
\caption{Performance (\%) comparison across methods on three datasets. Best results in bold, second-best underlined. ``*” denotes significant improvement over the second-best (IISAN-VS) at 0.05 level with paired T-test.}
\centering
\label{tab:eff_per}
\begin{tabular}{c|cc|cc|cc}
\hline
\multirow{2}{*}{Method} & \multicolumn{2}{c|}{Scientific} & \multicolumn{2}{c|}{Instrument} & \multicolumn{2}{c}{Office} \\
& H@10 & N@10 & H@10 & N@10 & H@10 & N@10 \\
\hline
FFT         & 6.62 & 4.06 & 8.76 & 6.76 & 6.30 & 4.58 \\
Adapter     & 6.61 & 3.91 & 8.82 & 6.83 & 6.65 & 4.85 \\
LoRA        & 6.62 & 4.09 & 8.43 & 6.64 & 6.55 & 4.78 \\
BitFit      & 6.37 & 3.76 & 8.65 & 6.71 & 6.78 & 4.87 \\
IISAN-VS    & \underline{6.83} & \underline{4.14} & \underline{9.06} & \underline{7.01} & \underline{6.80} & \underline{4.92} \\
\textbf{IISAN-VA} & \textbf{7.29$^*$} & \textbf{4.41$^*$} & \textbf{9.52$^*$} & \textbf{7.17$^*$} & \textbf{7.02$^*$} & \textbf{5.11$^*$} \\
\hline
\end{tabular}
\end{table}
To validate RQ1, which is whether incorporating a larger text encoder can enhance the performance of IISAN-Versa, this section evaluates the performance of an asymmetrical IISAN-Versa (IISAN-VA) using the BERT-large as text encoder compared to a symmetrical IISAN-Versa (IISAN-VS) with a BERT-base encoder. The specifics of the models are detailed in Table \ref{tab:model_specs}. In this section, we fix the image encoder as a ViT-base for a fair comparison, more image encoders will be further discussed in Section \ref{sec:scaling}. We also use FFT and popular EPEFT approaches (Adapter, LoRA, and BitFit) as baselines. FFT and EPEFT methods were applied only to the symmetrical model due to the high computational cost of EPEFT and FFT. From Table \ref{tab:eff_per}, IISAN-VS demonstrates superior performance over traditional full fine-tuning and embedded PEFT approaches (Adapter, LoRA, and BitFit), as evidenced by its effectiveness across the three Amazon datasets.

The asymmetrical IISAN-Versa (IISAN-VA) further enhances performance with the larger text encoder. It achieves a gain of +6.31\% in HR@10 and +6.12\% in NDCG@10 for the Scientific dataset, +4.83\% and +2.23\% for the Instrument dataset, and +3.13\% and +3.72\% for the Office dataset. These improvements underscore IISAN-VA's ability to effectively adapt to larger text encoders, leading to increased recommendation performance. This also demonstrates that compared to a BERT-base model, asymmetrical IISAN-Versa with a larger text encoder can generate a more powerful text representation for sequential recommendation tasks.

\textbf{(Answer to RQ1) IISAN-Versa can improve performance when equipped with a larger text encoder.} Furthermore, we conclude the overall findings as the following: (1) IISAN-VS demonstrates its effectiveness compared to traditional fine-tuning approaches, including FFT and EPEFT, as shown by its superior performance. (2) The IISAN-VA further enhances the performance of IISAN-VS, affirming the advantages of using a larger and more powerful text encoder over symmetrical IISAN-VS for generating powerful text representations in sequential recommendation tasks. 
Such a finding further highlights the IISAN-Versa's versatility in adapting both symmetrical and asymmetrical structures and their effectiveness.

\begin{table}[htbp]
    \centering
    \caption{IISAN‑Versa Encoders Scaling.}
    \renewcommand{\arraystretch}{0.9}
    \renewcommand\tabcolsep{12pt}
    \begin{tabular}{lrr} 
        \hline
        Model structure                                 &    HR@10   &  NDCG@10  \\
        \hline
        BERT‑base + ViT‑base                             & 0.0683 & 0.0414 \\
        BERT‑large + ViT‑base                            & 0.0729 & 0.0441 \\
        Gemma‑7b‑it + ViT‑base                           & 0.0732 & 0.0446 \\
        Gemma‑7b + ViT‑base                              & 0.0773 & 0.0475 \\
        Mistral‑7b + ViT‑base                            & 0.0753 & 0.0465 \\
        Llama2‑7b + ViT‑base                             & 0.0789 & 0.0481 \\
        Llama2‑chat‑7b + ViT‑base                        & 0.0749 & 0.0453 \\
        Llama3‑8b + ViT‑base                             & 0.0778 & 0.0463 \\
        Llama3‑8b‑it + ViT‑base                          & 0.0766 & 0.0474 \\
        Mixtral‑8×7b‑it + ViT‑base                       & 0.0798 & 0.0480 \\
        Mixtral‑8×7b + ViT‑base                          & 0.0781 & 0.0473 \\
        Llama3‑70B‑GPTQ‑it + ViT‑base                    & 0.0801 & 0.0478 \\
        Llama3‑70B‑GPTQ + ViT‑base                       & \textbf{0.0814} & \textbf{0.0494} \\
        \hline
        BERT‑base + ViT‑large                            & 0.0688 & 0.0425 \\
        BERT‑base + ViT‑huge                             & 0.0682 & 0.0415 \\
        BERT‑base + EVA‑ViT                              & 0.0746 & 0.0457 \\
        Llama3‑70B‑GPTQ + ViT‑large                      & 0.0756 & 0.0455 \\
        Llama3‑70B‑GPTQ + ViT‑huge                       & 0.0776 & 0.0464 \\
        Llama3‑70B‑GPTQ + EVA‑ViT                        & 0.0762 & 0.0469 \\
        \hline
        Llama3‑70B‑GPTQ + ViT‑tiny                       & 0.0780 & 0.0478 \\
        BERT‑tiny + EVA‑ViT                              & 0.0744 & 0.0449 \\
        \hline
    \end{tabular}
    \label{tab:scaling}
\end{table}

\begin{table}[h]
\centering
\caption{\textcolor{black}{Performance comparison of IISAN-Versa on a large dataset over 1M interactions. ``*” denotes that the improvements are significant at the level of 0.05 with paired T-test.}}
\label{tab:ninerec1m}
\renewcommand{\arraystretch}{1}
\begin{tabular}{cccc}
\hline
\multirow{2}{*}{\textcolor{black}{Class}} & \multirow{2}{*}{\textcolor{black}{Model}} & \multirow{2}{*}{\textcolor{black}{HR@10}} & \multirow{2}{*}{\textcolor{black}{NDCG@10}} \\
\\
\hline
\multirow{2}{*}{\textcolor{black}{MM VideoRec}}
& \textcolor{black}{Full finetuning} & \textcolor{black}{0.0098} & \textcolor{black}{0.0048} \\
& \textcolor{black}{Adapter} & \textcolor{black}{0.0097} & \textcolor{black}{0.0048}\\
\hline
\multirow{2}{*}{\textcolor{black}{IISAN-Versa}}
& \textcolor{black}{IISAN-VS (ours)} & \textcolor{black}{\underline{0.0105}} & \textcolor{black}{\underline{0.0052}} \\
& \textcolor{black}{IISAN-VA (ours)} & \textcolor{black}{\textbf{0.0111$^*$}} & \textcolor{black}{\textbf{0.0054$^*$}} \\
\hline
\end{tabular}
\vspace{-0.1in}
\end{table}

\vspace{-0.1in}
\subsection{Scaling Effect (RQ2)}
\label{sec:scaling}

In this section, we explore whether the IISAN-Versa exhibits a scaling effect. We perform experiments on the Scientific dataset, scaling up the text encoder across 14 models ranging from the basic BERT-base to the gigantic Llama-3-70b model. Since the available ViT models are ViT-base, ViT-large, and ViT-huge, we also performed scaling experiments on these image encoders. To better visualize the recommendation performance alongside the scale of model size, we categorize the models into four groups: PLM for the BERT model, Base LLM for models with 7 to 8 billion parameters, MoE LLM for the 8x7b Mixtral LLM, and 70b LLM.

As depicted in Table \ref{tab:scaling}, IISAN-Versa equipped with the largest scale of Llama3-70b-GPTQ model demonstrated the most superior performance.  The trend of these results is visualized in Figure \ref{fig:scaling}, which shows the overall trajectory is generally upward. Moreover, the average performance within each group significantly improves with each scale increase, supporting the hypothesis that IISAN-Versa follows a scaling effect for text encoders. This observation validates the IISAN-Versa framework's capability to leverage advancements from pre-trained foundation models in the NLP domain. 
We can observe that, with the exception of the Mixtral 8x7b model, most LLMs with instruction tuning typically do not outperform those without it. This reflects that the LLM without instruction tuning may generate a more powerful representation.

\begin{table}[h]
\centering
\caption{Performance comparison of IISAN-Versa on a multimodal public benchmark MicroLens. In this experiment, IISAN-VS and IISAN-VA adopt BERT-base and BERT-large respectively. ``*” denotes that the improvements of the best models compared with previous state-of-the-art GRU4Rec$_v$ are significant at the level of 0.05 with paired T-test.}
\label{tab:microlens}
\renewcommand{\arraystretch}{0.8}
\begin{tabular}{cccc}

\hline
\multirow{2}{*}{Class} & \multirow{2}{*}{Model} & \multirow{2}{*}{HR@10} & \multirow{2}{*}{NDCG@10} \\
\\
\hline
\multirow{4}{*}{IDRec (CF)} 
& DSSM \cite{huang2013learning}  & 0.0394 & 0.0193 \\
& LightGCN \cite{he2020lightgcn}   & 0.0372 & 0.0177 \\
& NFM \cite{he2017neural}  & 0.0313 & 0.0159 \\
& DeepFM \cite{guo2017deepfm} &   0.0350     &   0.0170 \\
\hline
\multirow{3}{*}{IDRec (SR)} 
& NextItNet \cite{yuan2018simple} & 0.0805 & 0.0442 \\
& GRU4Rec \cite{hidasi2015session} & 0.0782     & 0.0423 \\
& SASRec \cite{kang2018self} & 0.0909 & 0.0517 \\
\hline
\multirow{8}{*}{VIDRec (Frozen)} 
& YouTube\textsubscript{ID}  & 0.0461       & 0.0229 \\
& YouTube\textsubscript{ID+V} \cite{covington2016deep} &0.0392    &0.0188  \\
& MMGCN\textsubscript{ID} & 0.0141 & 0.0065  \\
& MMGCN\textsubscript{ID+V} \cite{wei2019mmgcn} & 0.0214 & 0.0103 \\
& GRCN\textsubscript{ID} & 0.0282 & 0.0131 \\
& GRCN\textsubscript{ID+V} \cite{wei2020graph}& 0.0306 & 0.0144 \\
& DSSM\textsubscript{ID+V}  & 0.0279 & 0.0137 \\
& SASRec\textsubscript{ID+V} & 0.0799 & 0.0415 \\
\hline
\multirow{3}{*}{VideoRec} 
& NexItNet\textsubscript{v}  \cite{ni2023content} & 0.0862 & 0.0466 \\
& GRU4Rec\textsubscript{v} \cite{ni2023content}  &0.0954 & 0.0517 \\
& SASRec\textsubscript{v}  \cite{ni2023content}  & 0.0948 & 0.0515 \\
\hline
MM VideoRec
& Full finetuning & 0.0926 & 0.0493 \\
\hline
\multirow{2}{*}{IISAN-Versa}
& IISAN-VS (ours) & \underline{0.0960} & \underline{0.0530} \\
& IISAN-VA (ours) & \textbf{0.0964$^*$} & \textbf{0.0533$^*$} \\
\hline
\end{tabular}
\vspace{-0.2in}
\end{table}

As shown in Table \ref{tab:scaling}, replacing ViT-base with ViT-large or ViT-huge yields only minor improvements, and in some cases even reduces performance—especially when using strong text encoders like Llama3-70B-GPTQ. To explore further, we include EVA-CLIP-18B which contains one of the largest publicly available vision transformers. While it significantly boosts performance when paired with lightweight text models, its impact is limited when the text encoder is already strong. This suggests that once strong textual representations are in place, the value of visual features diminishes—a trend aligned with prior findings \cite{fu2024exploring,yuan2023go} and supported by our ablation studies showing the dominant role of text in multimodal recommendation. To further strengthen our analysis, we include extreme architectural asymmetry—where the text encoder is significantly larger or smaller than the vision encoder—demonstrating the robustness of our IISAN-Versa.

Beyond model-wise scaling, we evaluate IISAN-Versa on the large-scale NINERec dataset (over 1M interactions) as shown in Table~\ref{tab:ninerec1m}. IISAN-VS and IISAN-VA achieve competitive performance over FFT and Adapter, highlighting the practical effectiveness of our approach.

\textbf{(Answer to RQ2) IISAN-Versa complies with the scaling effect of text encoders, confirming its potential to inherit the benefits of advanced pre-trained text transformer models.} This also indicates that researchers can explore LLMs not only as powerful text generators but also as effective text encoders in future works, thus reaching more comprehensive representations for recommender systems. 
\textcolor{black}{Further analysis shows that a large vision encoder, such as EVA-ViT, can effectively compensate when the text encoder is relatively weak, leading to noticeable performance gains. However, its impact diminishes once the text encoder is already strong, suggesting that visual enhancements offer limited benefit in such cases. Therefore, scaling the vision encoder is most beneficial in asymmetric setups with lightweight text models, but less effective when paired with high-capacity text encoders.}

\begin{figure}[h]
  \centering
  \begin{adjustbox}{minipage=[t]{0.57\linewidth},valign=t}
    \centering
    \includegraphics[width=\linewidth]{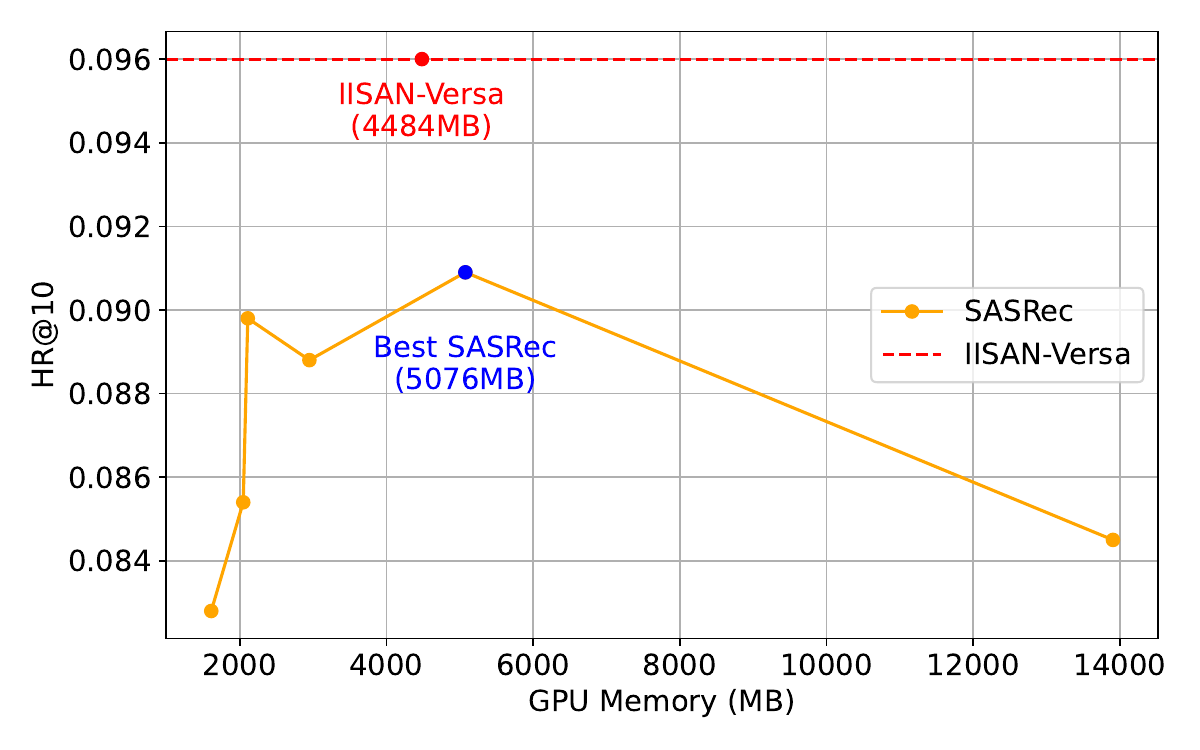}
    \caption{\textcolor{black}{Comparison of GPU Memory Usage and Performance Between SASRec and IISAN-Versa.}}
    \label{fig:compare_id_iisan}
  \end{adjustbox}
  \hfill
  \begin{adjustbox}{minipage=[t]{0.40\linewidth},valign=t}
    \centering
    \captionof{table}{Efficiency comparison with optimal SASRec and IISAN-Versa. GM represents GPU memory (GB), TT represents Training time (seconds) per epoch.}
    \renewcommand\tabcolsep{1pt}
    \begin{tabular}{lcc}
      \toprule
      \textbf{Model} & \textbf{GM} & \textbf{TT} \\
      \midrule
      IISAN-Versa & 4.38 & 51 \\
      SASRec & 4.96 & 9 \\
      \bottomrule
    \end{tabular}
    \label{tab:compare_id_iisan}
  \end{adjustbox}
\end{figure}

\subsection{Performance on Public Benchmark (RQ3)}
\label{sec:microlens}
To further evaluate IISAN-Versa, we compare it with the public benchmark MicroLens, as detailed in Table \ref{tab:microlens}.  This public benchmark \cite{ni2023content} includes four main types of baselines: sequential recommendation (IDRec (SR)), collaborative filtering (IDRec (CF)), and ID-based recommendation with features generated from frozen pre-trained video encoder (VIDRec (Frozen Encoder)) or end-to-end (E2E) learning encoders (VideoRec (E2E learning)). The previous state-of-the-art approach, GRU4Rec$_V$ \cite{ni2023content}, utilizes GRU4Rec as the sequential encoder with a tunable video encoder VideoMAE to encode video content in an E2E manner directly. To make sure a fair comparison we directly adopt the data from the original benchmark~\cite{ni2023content}. We also perform a comprehensive comparison by including a multimodal full fine-tuning of the multimodal VideoRec framework, as applied in \cite{fu2024iisan}.

\begin{figure}
  \centering
   \includegraphics[width=\linewidth]{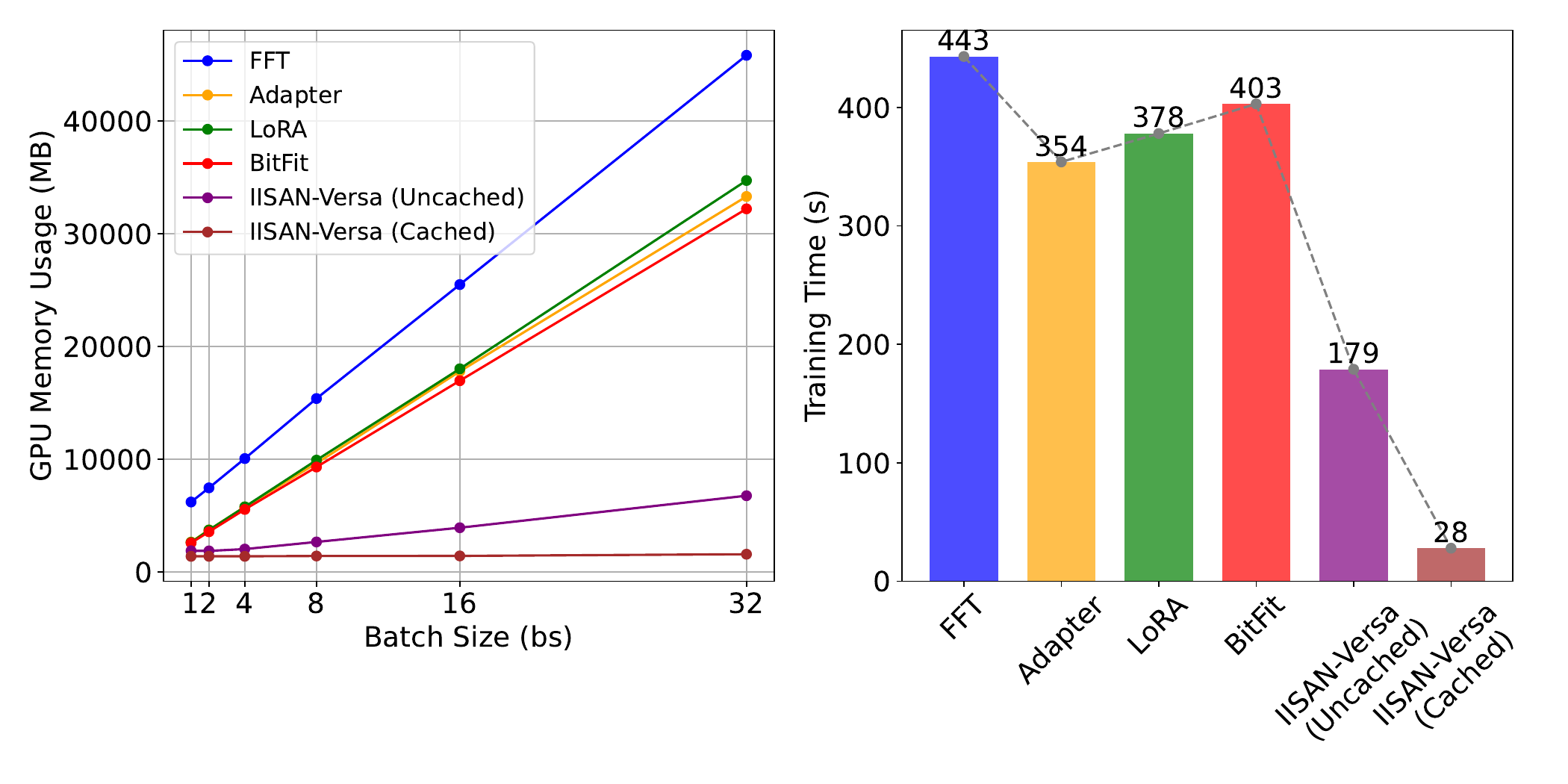}
     \vspace{-2em}
  \caption{GPU Memory and Training time comparison. The left figure illustrates the training time per epoch for each adaptation method. The figure on the right demonstrates the GPU Memory usage for different adaptation methods according to the different number of batch sizes. 
  }
  \vspace{-0.1in}
    \label{fig:efficency} 
\end{figure}

Figure \ref{fig:compare_id_iisan} and Table~\ref{tab:compare_id_iisan} presents a comparison between our proposed model and the widely adopted ID-based baseline SASRec, focusing on both effectiveness and computational efficiency. \textcolor{black}{Although SASRec demonstrates strong performance in sequential recommendation, our model consistently achieves higher accuracy across all metrics while requiring less GPU memory. While multimodal processing inherently introduces additional training time, the combination of reduced memory consumption and improved accuracy underscores the practical potential of our approach. These results affirm the real-world viability of our method, particularly in deployment scenarios where both performance and GPU memory are essential.}

\begin{figure}
  \centering
   \includegraphics[width=0.7\linewidth]{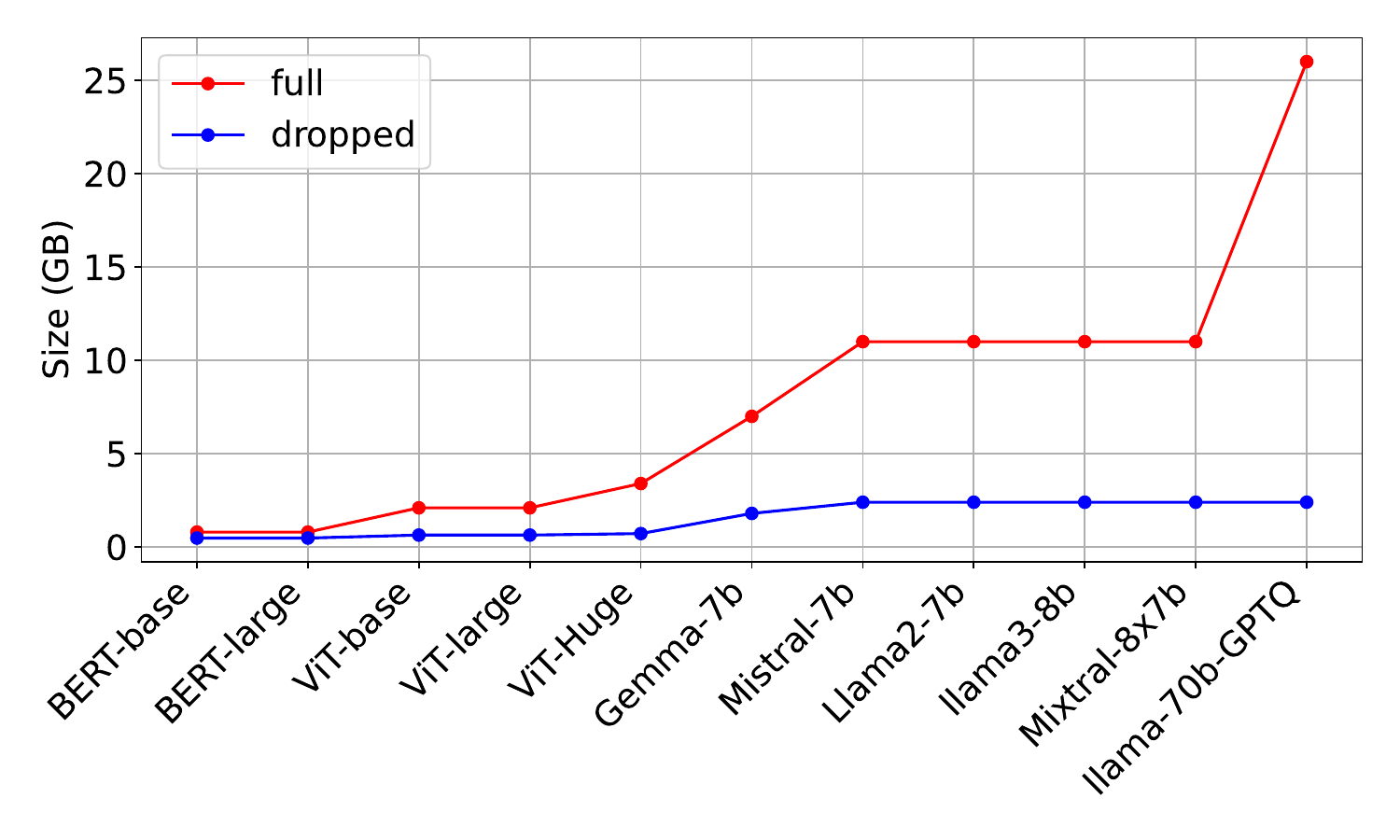}
       \vspace{-0.15in}
  \caption{On-Disk Storage Requirements for Caching Strategy, by Model Size. The x-axis represents the increasing model size, with the Scientific dataset, comprising 20,314 embedding files to be cached.
  }

    \label{fig:storage_efficency} 
    \vspace{-0.15in}
\end{figure}

\begin{figure}
  \centering
   \includegraphics[width=0.9\linewidth]{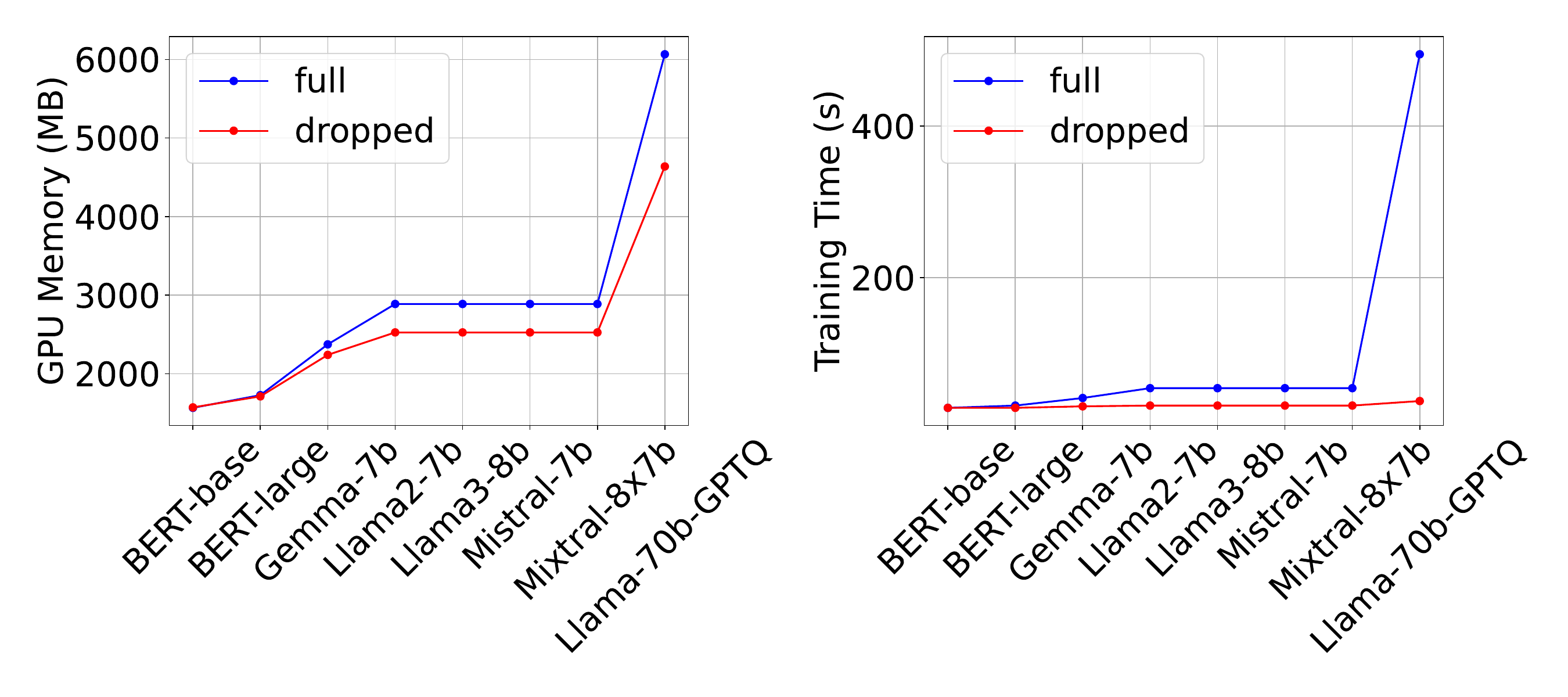}
   \vspace{-0.2in}
  \caption{GPU memory and Training time usage for IISAN-Versa with different text encoders. 'Full' represents using the entire cached embedding, whereas 'dropped' indicates retaining only the useful cached embeddings. 
  }
    \label{fig:va_efficency} 
    \vspace{-0.1in}
\end{figure}

As shown in Table \ref{tab:microlens}, both IISAN-Versa approaches outperform existing approaches on this dataset. Notably, IISAN-Versa can significantly reduce training costs compared to VideoRec, which requires over four times more GPU resources than even full fine-tuning of the multimodal paradigm. This is due to the high computational cost of video encoders that process multiple frames during training. Our approach, with its caching strategy, results in IISAN-Versa using vastly smaller GPU memory. To further validate IISAN-Versa on MicroLens, we also incorporate the standard full fine-tuning method, which underperforms the IISAN-Versa. Within the IISAN-Versa paradigm, IISAN-VA demonstrates better performance compared to IISAN-VS, which is consistent with our previous findings.

\textbf{(Answer to RQ3) IISAN-Versa surpasses state-of-the-art performance on the public benchmark MicroLens dataset.}

\subsection{Efficiency Analysis (RQ4)}
\label{sec:eff_experiment}
In Section \ref{sec:eff_analysis}, we provided a comprehensive analysis of various adaptation techniques. This section presents a detailed evaluation of experimental efficiency, focusing on two key aspects: GPU memory usage and training speed across full fine-tuning, EPEFT, and DPEFT methods. We provide details on the time taken per epoch and the GPU memory consumption for each method. Additionally, we report the on-disk memory usage for the cached hidden states of each pretrained model

Figure \ref{fig:efficency} illustrates the variation in GPU memory usage as the batch size increases from \{1, 2, 4, 8, 16, 32\}. We utilize the BERT-base and ViT-base as the fixed text and vision encoders, respectively. Further studies on the efficiency of various encoder sizes will be discussed in subsequent sections. The results show that IISAN-Versa significantly reduces GPU memory usage as the batch size grows, indicating that the decoupled PEFT approach employed by IISAN-Versa effectively minimizes activations during training. Since activations are directly proportional to batch size, this reduction is particularly noteworthy. The EPEFT methods experience a substantial increase as the batch size scales up, making their practical memory savings less impactful compared to the IISAN-Versa. Additionally, IISAN-Versa also exhibits a significant improvement in training speed.

With the caching strategy, we essentially trade on-disk memory (cached for hidden states) for enhanced training efficiency in terms of both time and GPU memory. To demonstrate practical storage requirements, we employ the Scientific dataset as an example and exclude the model with instruction tuning, given that its hidden state shapes are identical to those of the model without instruction tuning. As shown in Figure \ref{fig:storage_efficency}, the on-disk storage requirement increases with the model size. The maximum storage required for IISAN-Versa (Cached) is for the Llama-70b-GPTQ model, which needs 26 GB of memory for the entire Scientific dataset, which is not excessively large for a cluster machine. If the cluster does not have sufficient on-disk memory, the storage can be reduced by the number of layers of the model. For example, after applying our k-group layerdrop strategies mentioned in Section \ref{sec:layerdrop}, the Llama-70b-GPTQ model requires only six layers and 2.4 GB of storage.\footnote{The storage of hidden states from Llama-70b-GPTQ is identical to that of Mistral, Llama2, etc., because it uses float16 as the datatype, whereas other base LLMs use float32. Additionally, the base LLMs have half the dimension size of the 70B LLM and the same number of layers after removing the unnecessary layers. Consequently, their storage requirements are the same.}

Furthermore, the efficiency of IISAN-Versa (Cached) with different sizes of text encoder is demonstrated in Figure \ref{fig:va_efficency}. The GPU memory usage increases with the model size, but it can be slightly reduced by offloading hidden states of unnecessary layers to GPU memory. Furthermore, it is worth noting that Llama-70b-GPTQ significantly increases the training cost when we directly use its full embeddings. Given their substantial size, we have observed that loading and processing these models on the CPU becomes the primary bottleneck rather than the GPU computation. Consequently, by reducing the embeddings to only six essential layers, we have achieved a significant reduction in training time.

\begin{table}
  \caption{Ablation study for IISAN-Versa and IISAN-VA on Scientific Dataset. IISAN-Versa mainly contains four key components: LayerDrop, Modality Gate, and Intra- and Inter-modal towers. \textcolor{black}{``}Frozen Backbone" refers to keeping the modality backbone frozen, without applying any adaptation methods.}
  \label{tab:abl_iisan}
  \renewcommand{\arraystretch}{0.9}
  \begin{tabular}{c | c c | c c}
    \hline
    \multirow{2}{*}{Method}&\multicolumn{2}{c|}{IISAN-VS}&\multicolumn{2}{c}{IISAN-VA}\\
    \cline{2-5}
    &HR@10&NDCG@10&HR@10&NDCG@10\\
    \hline
    -w/o LayerDrop&6.73 & 4.04& 6.85& 4.11 \\
    \cdashline{2-5}
    -w/o Modality Gate&6.58 &3.89&7.08 &4.30\\
    \cdashline{2-5}
    Frozen Backbone &6.00 & 3.53 &5.81 &3.37 \\
    -w/o Inter-modality&6.38 &3.89 &6.96 & 4.21\\
    -w/o Intra-modality&6.41 & 3.83 &7.12   &4.43\\
    \cdashline{2-5}
    IISAN-Versa&6.83 & 4.14&7.29 & 4.41\\
  \hline
\end{tabular}
\end{table}

\begin{figure}
    \centering
    \includegraphics[width=\linewidth]{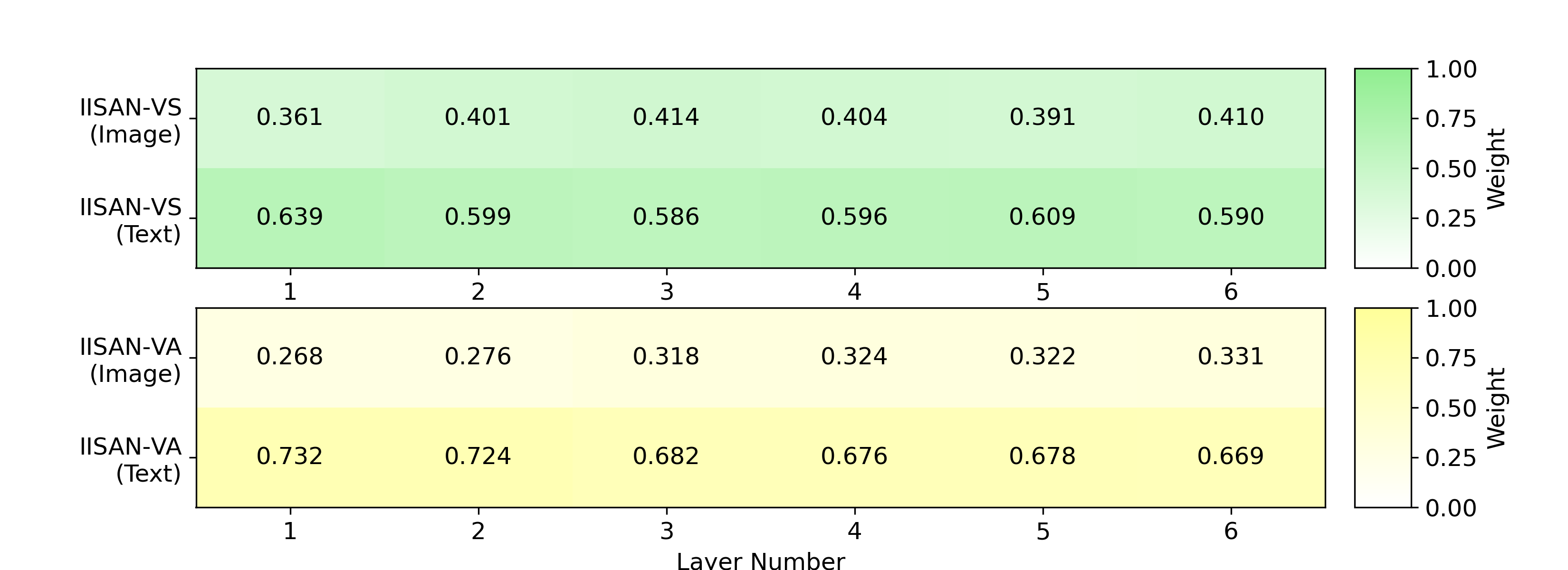}
    \caption{Visualization of inter-modality gates in IISAN-Versa. Darker colors indicate higher gate weights.}
    \label{fig:visualization_heatplot}
\end{figure}

\begin{table}
\centering
  \caption{LayerDrop in symmetrical IISAN-Versa on Scientific Dataset. The number of blocks in the Method column represents that keep this number of blocks and drop the others.}
  \label{tab:layerdrop}
  \renewcommand{\arraystretch}{0.9}
  \scalebox{0.9}{
  \begin{tabular}{c | c | c  c}
    \hline
    \multirow{2}{*}{Method} & \multirow{2}{*}{Number of blocks} & \multirow{2}{*}{HR@10} & \multirow{2}{*}{NDCG@10} \\
    & & & \\
    \hline
    \multirow{5}{*}{IISAN-VS} & 2 blocks & 6.46 & 3.88 \\
    & 3 blocks & 6.79 & 4.12  \\
    & 4 blocks & 6.57 & 3.98 \\
    & \textbf{6 blocks} & \textbf{6.83} & \textbf{4.14} \\
    & 12 blocks & 6.73 & 4.04 \\
  \hline
  \end{tabular}
  }
  \vspace{-0.1in}
\end{table}

\begin{table}[h!]
\centering
\caption{Ablation Study for blocks of asymmetrical IISAN-Versa. n(\textcolor{black}{uniform\_}all) refers to selecting n blocks evenly distributed across all blocks. \textcolor{black}{``}Top" and \textcolor{black}{``}down" indicate that the blocks are drawn from either the upper half or the lower half, respectively.}\label{tab:layerdrop_va}
\renewcommand{\arraystretch}{0.9}
\scalebox{0.9}{
\begin{tabular}{cccc}
\hline
\multirow{2}{*}{Method} & \multirow{2}{*}{Configuration} & \multirow{2}{*}{HR@10} & \multirow{2}{*}{NDCG@10} \\
\\
\hline
24 + 12 & Full & 6.85 & 4.11 \\
\hline
\multirow{4}{*}{12+12}
& 12(\textcolor{black}{uniform\_}all)+12(all) & 6.96 & 4.29 \\
& 12(top)+12(all) & 6.96 & 4.27 \\
& 12(down)+12(all) & 6.82 & 4.21 \\
\hline
& Mean (12+12) & 6.91 & 4.26 \\
\hline
\multirow{10}{*}{12+6}
& 12(\textcolor{black}{uniform\_}all)+6(\textcolor{black}{uniform\_}all) & 7.08 & 4.44 \\
& 12(\textcolor{black}{uniform\_}all)+6(top) & 6.80 & 4.19 \\
& 12(\textcolor{black}{uniform\_}all)+6(down) & 7.15 & 4.29 \\
& 12(top)+6(top) & 6.94 & 4.22 \\
& 12(top)+6(\textcolor{black}{uniform\_}all) & 7.25 & 4.37 \\
& 12(top)+6(down) & 6.67 & 4.04 \\
& 12(down)+6(top) & 7.14 & 4.30 \\
& 12(down)+6(\textcolor{black}{uniform\_}all) & 6.84 & 4.24 \\
& 12(down)+6(down) & 7.26 & 4.43 \\
\hline
& Mean (12+6) & 7.01 & 4.28 \\
\hline
\multirow{10}{*}{6+6}
& 6(\textcolor{black}{uniform\_}all)+6(\textcolor{black}{uniform\_}all) & 7.29 & 4.41 \\
& 6(\textcolor{black}{uniform\_}all)+6(down) & 7.21 & 4.37 \\
& 6(\textcolor{black}{uniform\_}all)+6(top) & 6.80 & 4.19 \\
& 6(\textcolor{black}{uniform\_}top12)+6(\textcolor{black}{uniform\_}all) & 6.96 & 4.24 \\
& 6(\textcolor{black}{uniform\_}top12)+6(top) & 6.95 & 4.27 \\
& 6(\textcolor{black}{uniform\_}top12)+6(down) & 6.97 & 4.25 \\
& 6(\textcolor{black}{uniform\_}down12)+6(\textcolor{black}{uniform\_}all) & 7.29 & 4.44 \\
& 6(\textcolor{black}{uniform\_}down12)+6(top) & 6.87 & 4.26 \\
& 6(\textcolor{black}{uniform\_}down12)+6(down) & 7.35 & 4.51 \\
\hline
& Mean (6+6) & \textbf{7.08} & \textbf{4.33} \\
\hline
\end{tabular}
}
\vspace{-0.1in}
\end{table}

\begin{table}[h]
\centering
\renewcommand{\arraystretch}{0.9}
\caption{Performance on Multimodal Text.}
\label{tab:multimodal_text}
\scalebox{0.9}{
\begin{tabular}{cccc}
\hline
\multirow{2}{*}{Category} & \multirow{2}{*}{Method} & \multirow{2}{*}{HR@10} & \multirow{2}{*}{NDCG@10} \\
\\
\hline
\multirow{3}{*}{FFT}
& Title & 0.0916 & 0.0490 \\
& Video Caption & 0.0878 & 0.0470 \\
& Cover Caption & 0.0915 & 0.0493 \\
\hline
\multirow{3}{*}{Same input}
& Title x2 & 0.0917 & 0.0507 \\
& Video Caption x2 & 0.0866 & 0.0476 \\
& Cover Caption x2 & 0.0892 & 0.0491 \\

\hline
\multirow{3}{*}{Multimodal Text}
& Title-Video Caption & \underline{0.0951} & \underline{0.0517} \\
& Title-Cover Caption & 0.0934 & 0.0511 \\
& Video-Cover Caption & 0.0899 & 0.0497 \\
\hline
Multimodal & IISAN-Versa(Text-Image) & \textbf{0.0960} & \textbf{0.0530} \\
\hline
\end{tabular}
}
\vspace{-0.2in}
\end{table}

\textbf{(Answer to RQ4) IISAN-Versa demonstrates a marked efficiency improvement over EPEFT approaches in both training time and GPU memory.} Furthermore, while the caching strategy incurs additional on-disk memory costs, these are generally manageable in most cases. Moreover, these costs can be further reduced by selectively caching only the useful layers after layer drop. Utilizing the useful layers can also significantly enhance efficiency when dealing with large language model (LLM) encoders.

\subsection{Ablation Studies (RQ5)}
\label{sec:ablation}
In this section, we first demonstrate the importance of the module's layer dropping for both symmetrical and asymmetrical IISAN-Versa.  The BERT-large is applied as the text encoder for IISAN-VA's ablation study. We extensively explore the blocks required for asymmetrical IISAN-Versa. Specifically, we experiment with the combinations of blocks, ranging from \{24+12, 12+12, 12+6, 6+6\}, with the former value in each combination indicating the blocks for 24-layer BERT-large and the latter representing the blocks for 12-layer ViT-base.

\textbf{(1) Modality Selection.} Table \ref{tab:abl_iisan} highlights that using separate intra-modal SAN (Line 4) or inter-modal SAN (Line 5) clearly reduces recommendation effectiveness. However, compared to the pre-trained backbones (Line 3) with frozen layers (where only the sequential encoder is trained), both intra-modal SAN and inter-modal SAN can significantly improve performance, performing best when used simultaneously. This reflects that intra- and inter-modal SAN can enhance feature adaptation within each modality and improve inter-modal feature interaction, such argument holds for both IISAN-VS and IISAN-VA.

\textbf{(2) Gated Fusion.} To deepen our understanding of the differences in information usage between textual and visual modalities, we analyzed the gate weights at the optimal checkpoint for both symmetrical and asymmetrical IISAN-Versa. \textcolor{black}{We visualize the gate values in Figure \ref{fig:visualization_heatplot}. Since text gate values are consistently above 0.5, all models primarily rely on text. Visual gate values stay between 0.26 and 0.42, with IISAN-VA depending on text even more than IISAN-VS.}  This trend suggests a predominant reliance on textual modality within our multimodal approach.

\textbf{(3) LayerDrop Strategies for IISAN-Versa.} IISAN-VS's layers exhibit a degree of redundancy, allowing LayerDrop to enhance performance effectively. First, when we remove LayerDrop, both HR@10 and NDCG@10 decrease by 0.1\%, demonstrating LayerDrop's effectiveness. Second, we study the effect of different LayerDrop strategies, as described in Table \ref{tab:layerdrop}. We adopt the even-numbered $(2, 4, 6, \ldots, 12)$ transformer blocks (6 blocks), skipping the odd-numbered layers, which achieves the best performance-efficiency balance. In terms of IISAN-VA, the performance drop is even more significant when LayerDrop is completely removed, indicating that the redundancy is more prominent than in IISAN-VS. To explore how many blocks should be used for IISAN-VA, we perform extensive experiments on the combination of the layers and compare their mean values (see Table~\ref{tab:layerdrop_va}).  Due to the complexity of the combination, we study the even selection for layer group of \{top, down, all\} for the 6+6 combination. When the numbers are asymmetrical such as ``24+12'' and ``12+6'', we adopt the top hidden states to align with the vision tower for Inter-SAN. The intra-modality stays the same. Extensive experiments show that the ``6+6'' structure has the best overall mean performance among all structures. This further supports the point that IISAN-VA has more layer redundancy compared to IISAN-VS, due to its greater number of incorporated layers.\footnote{Exhaustively testing all layer combinations, even for smaller models like IISAN-VA (BERT-large + ViT-base), would require about 68 billion evaluations. Instead, we aim to find a general strategy for larger models. We adopt the 6(\textcolor{black}{uniform\_}all) + 6(\textcolor{black}{uniform\_}all) method as the default for IISAN-VA, which, while not the best, performs above average among 6+6 combinations in Table \ref{tab:layerdrop_va}, offering convenience and generality for asymmetrical structures.}

\textbf{(Answer to RQ5)} In this section, we conclude with three key findings regarding IISAN-Versa's components: (1) Optimal performance is achieved through the integration of both intra- and inter-modal SANs for both symmetrical and asymmetrical structures. (2) The best performance is achieved by dropping half of the SANBs in the intra- and inter-modal SANs for IISAN-VS. For IISAN-VA, the best performance comes from using the least number of blocks, transforming the number of layers to be symmetrical to the vision encoders.
(3) Text-image interaction is effective, but the text modality plays a more crucial role in the recommendation, where inter-modal SAN can effectively maintain the dominance of text modality while integrating image information.

\vspace{-0.1in}

\subsection{Performance on Multimodal Text (RQ6)}
\label{sec:multimodal_text}

Multimodal text encompasses different types of text, such as captions extracted from raw images and videos. These settings for their multimodal recommendation systems when they lack the computational resources to process images and videos \cite{zhang2023multimodal} or when the datasets do not include images, which is often the case. Therefore, we aim to explore how IISAN-Versa performs in this scenario. The MicroLens dataset \cite{ni2023content} is used for experimentation due to its rich modality information, we can extract captions from both cover images and video content. In the multimodal text experiments, we use IISAN-Versa to encode two combinations of text: raw text (title) and generated caption text (video/cover caption), with each tower encoding a specific text modality. To isolate the impact of the modality itself and minimize the influence of the text encoder's power, we use an IISAN-Versa model with two BERT-base models to evaluate its performance. 

The results, shown in Table \ref{tab:multimodal_text}, demonstrate that IISAN-Versa effectively adapts to multimodal text by outperforming both title-based and caption-based approaches. This superior performance over these three baselines highlights IISAN-Versa's versatility. Additionally, the combination of raw text—title and video caption—proves to be the most informative. By learning from this combination of these texts, we achieve the best overall performance in multimodal text scenarios. However, it is important to note that no combination of multimodal text outperforms the use of titles with raw cover images\footnote{Note that involving raw video and audio are beyond the scope of this paper.}, indicating that raw modality remains crucial in the multimodal representation of sequential recommendations.

\textbf{(Answer to RQ6) (1) IISAN-Versa is a viable option for multimodal text and effectively incorporates multiple types of text information.} (2) According to the experiment, titles and video captions are the most important sources for IISAN-Versa. (3) Despite the effectiveness of incorporating multimodal text compared to using only one type of text, we find that IISAN-Versa with raw multimodality achieves the best performance.

\vspace{-0.1in}
\subsection{Hyper-parameter Sensitivity Analysis (RQ7)}
\label{sec:hyper-parameter}

In this section, we address RQ7 by conducting a sensitivity analysis on two key hyperparameters—learning rate and embedding dimension—following the setup in \cite{fu2024exploring}, to provide a more comprehensive empirical evaluation of PEFT-based modality recommendation.

\begin{figure}
  \centering
   \includegraphics[width=\linewidth]{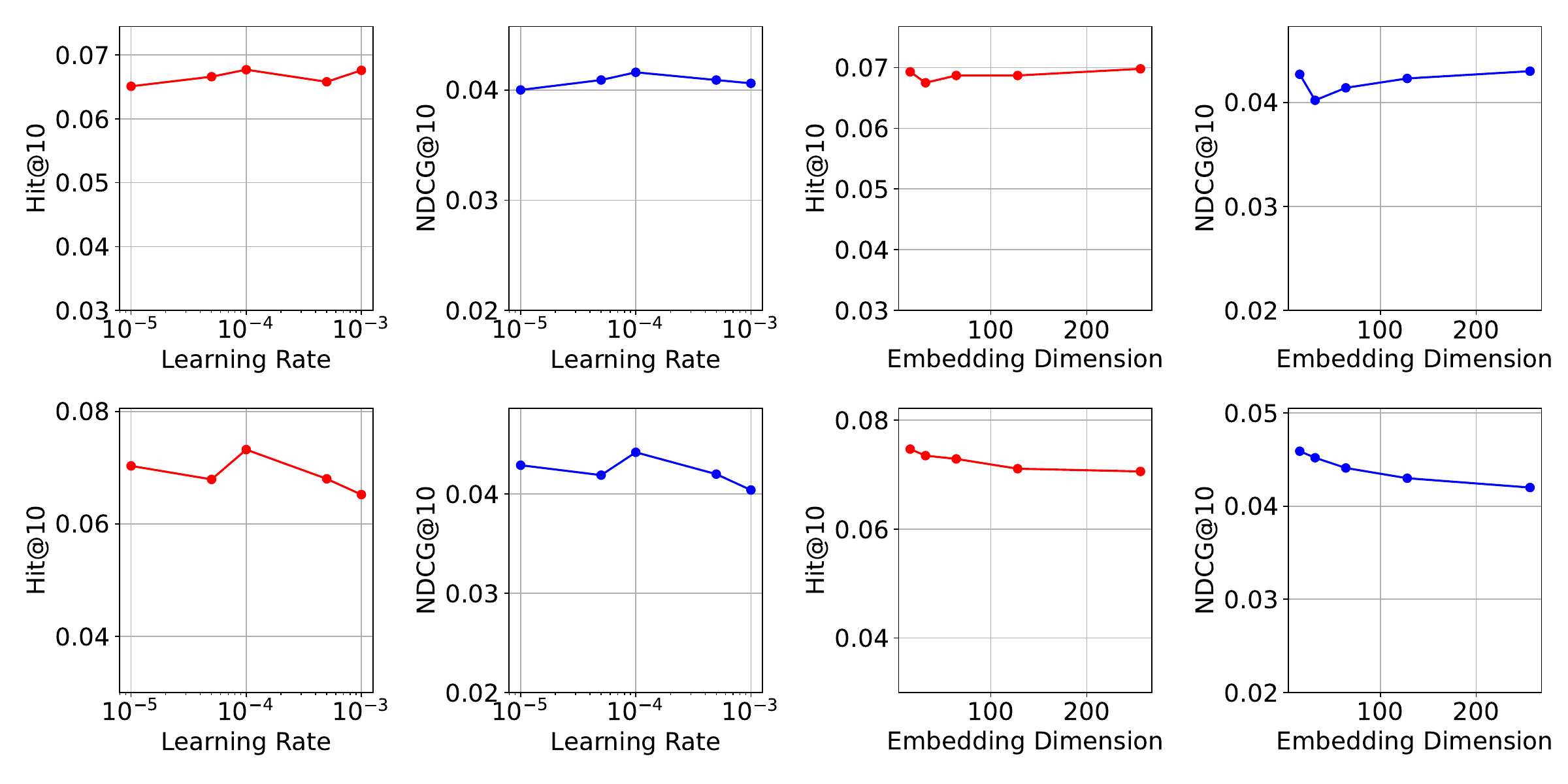}
  \caption{\textcolor{black}{Learning rate and Embedding Dimension of IISAN-Versa. The first row represents the performance of IISAN-VS on the Scientific dataset whereas the second row is for IISAN-VA.}
  }
  \vspace{-0.2in}
    \label{fig:hyper} 
\end{figure}

As shown in Figure~\ref{fig:hyper}, both IISAN-Versa variants demonstrate robust performance across a range of hyperparameter settings. IISAN-VS maintains stable performance regardless of the learning rate or embedding dimension. IISAN-VA also remains stable overall, but its performance slightly degrades when either the embedding dimension or the learning rate is set too high. The best performance is achieved when both hyperparameters are set to relatively small values; notably, smaller embedding dimensions can enhance efficiency.

\textbf{(Answer to RQ7)} Both IISAN-VS and IISAN-VA exhibit robustness to hyperparameter variations. However, IISAN-VA requires more conservative tuning, with smaller embedding dimensions and moderate learning rates yielding improved performance and efficiency.

\section{Conclusion and Future Work}
This paper introduces the new IISAN-Versa, an extension of the IISAN framework, which offers a versatile and efficient architecture for multimodal sequential recommendation tasks. IISAN-Versa overcomes the previous limitation of being applicable only to large asymmetrical MFMs, demonstrating a clear scaling effect where larger text models consistently enhance recommendation performance. Additionally, IISAN-Versa proves effective in handling diverse multimodal text scenarios, significantly broadening its applicability. 

Future research should aim to overcome current limitations by incorporating additional modalities such as audio, raw video, or graph information. \textcolor{black}{Furthermore, exploring the potential of adapting the IISAN-Versa framework to tasks such as Visual Question Answering (VQA) and video retrieval is a promising direction, as these tasks commonly adopt dual-tower architectures where one tower encodes the visual modality and the other the textual input. IISAN-Versa naturally fits into this paradigm, with its modality-specific intra-SANs and Inter-SAN facilitating effective interaction between different modalities. Extending to additional modalities like audio or video can be easily achieved by adding more intra-SANs for each modality and including them in the Inter-SAN computation, enabling flexible multimodal fusion while maintaining adaptation efficiency.} Another promising research direction involves optimizing training efficiency by reducing the number of trainable parameters, particularly in large text encoders like Llama-70B, given their extensive embedding dimensions.
\appendices

\ifCLASSOPTIONcaptionsoff
  \newpage
\fi



%
\bibliographystyle{IEEEtran}
\bibliography{sample-base}
%
\vskip -0.4in
\begin{IEEEbiography}[{\includegraphics[width=1in,height=1.25in,clip,keepaspectratio]{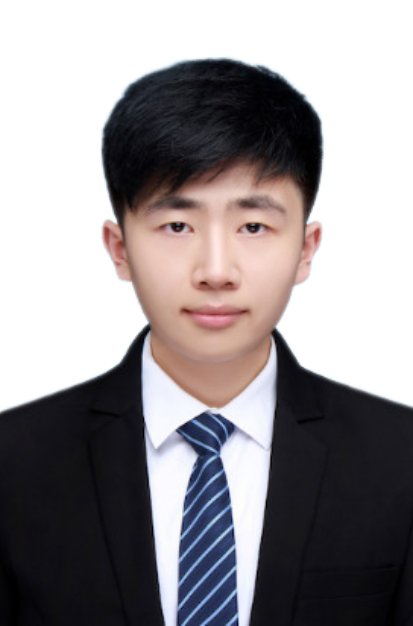}}]{Junchen Fu} is a third-year PhD candidate under the
supervision of Prof. Joemon Jose at the School of Computing Science, University of Glasgow. He has contributed to several leading AI conferences and
journals, including ICML, SIGIR, WWW, WSDM,
CIKM, MM, CVPR, and TPAMI. He also serves as
the program committee member and invited reviewer for premier conferences and journals, such as ICLR, SIGIR, KDD, WSDM, CIKM, AAAI, MM, TKDE, TOIS, TMM, and TORS. His research primarily focuses on enhancing the efficiency of adapting
multimodal foundation models for recommendation.
\end{IEEEbiography}
\vskip -0.4in
\begin{IEEEbiography}[{\includegraphics[width=1in,height=1.25in,clip,keepaspectratio]{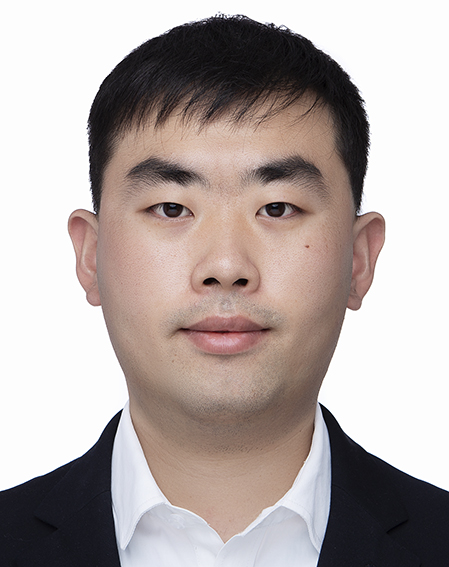}}]{Xuri Ge} is now a tenure-track assistant professor in the school of artificial intelligence, Shandong University. He earned his PhD at the University of Glasgow (UK) and received M.S. degree from Xiamen University (China). His current research interests include computer vision, multimodal representation learning, and information retrieval. He has contributed to several leading conferences and journals, including ICML, NeurIPS, SIGIR, ACM MM, CIKM, WSDM, ACM TIST, and IP\&M, etc. He serves as the PC member and reviewer for conferences and journals, such as NeurIPS, ICLR, TKDE, TOIS, IJCV, etc.
\end{IEEEbiography}
\vskip -0.4in
\begin{IEEEbiography}[{\includegraphics[width=1in,height=1.25in,clip,keepaspectratio]{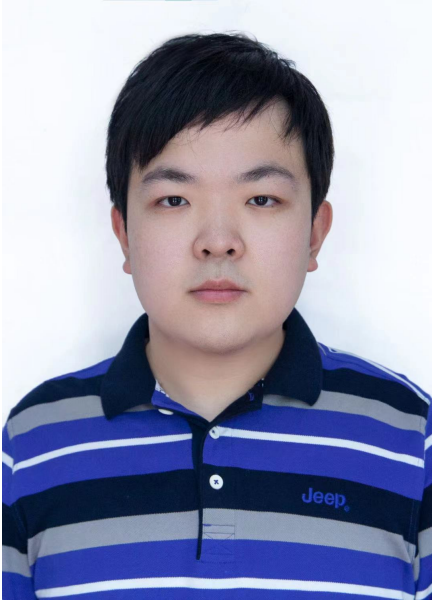}}]{Xin Xin} is now a tenure-track associate professor in the school of computer science and technology, Shandong University. His research interests include machine learning and reinforcement learning for recommender systems and information retrieval. He has published more than 30 papers in top-ranking conferences, including SIGIR, IJCAI, ACL, WSDM, etc. He also serves as the program committee member and invited reviewer for tier-1 conferences and journals, such as SIGIR, IJCAI, WSDM, ACL.
\end{IEEEbiography}

\vskip -0.4in
\begin{IEEEbiography}[{\includegraphics[width=1in,height=1.25in,clip,keepaspectratio]{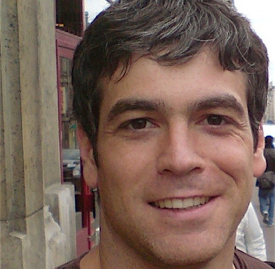}}]{Alexandros Karatzoglou} is a Principal Applied Scientist at Amazon. His contributions, such as GRU4Rec and kernlab, have significantly influenced and advanced future research in machine learning and recommender systems. His work has garnered over 17,000 Google Scholar citations, with an h-index of 42. Before joining Amazon, he was a Staff Research Scientist at Google DeepMind. He also served as Director of Telefonica Research in Barcelona and holds a PhD from the Vienna University of Technology.
\end{IEEEbiography}

\vskip -0.4in
\begin{IEEEbiography}[{\includegraphics[width=1in,height=1.25in,clip,keepaspectratio]{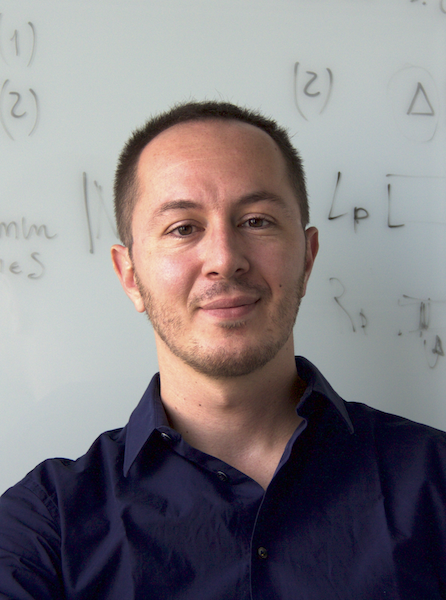}}]{Ioannis Arapakis} is a Principal Research Scientist at Telefónica Research, specializing in behavior interpretation algorithms for user modeling in both offline and online contexts, with a focus on web search. He holds a Ph.D. in Information Retrieval from the University of Glasgow, and an M.Sc. in Information Technology from the Royal Institute of Technology. Previously, he was a Research Scientist at Yahoo Labs, where he worked on data mining, IR, HCI, and multimedia mining projects. 
\end{IEEEbiography}
\vskip -0.4in
\begin{IEEEbiography}[{\includegraphics[width=1in,height=1.25in,clip,keepaspectratio]{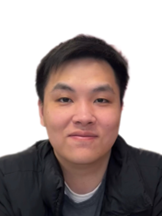}}]{Kaiwen Zheng} is a third-year Ph.D. candidate in
the School of Computing Science at the University of Glasgow, under the supervision of Prof. Joemon Jose. His research primarily focuses on multimodal expression recognition. He has published in leading AI conferences and journals, including ICML, WWW, and ICME, and serves as an invited reviewer for premier venues such as AAAI, IEEE Transactions on Affective Computing (TAFFC), ICME, and CIKM.
\end{IEEEbiography}
\vskip -0.4in
\begin{IEEEbiography}[{\includegraphics[width=1in,height=1.25in,clip,keepaspectratio]{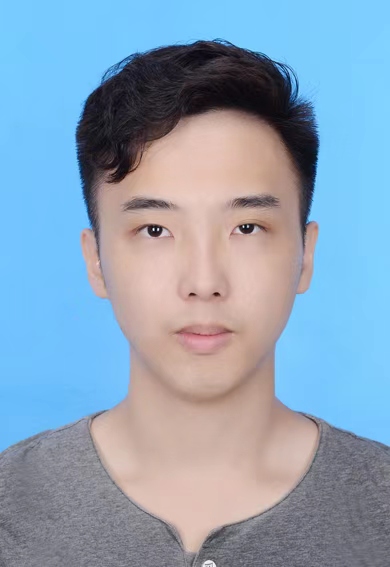}}]{Yongxin Ni} is currently a research assistant at University of Science and Technology of China. He received the B.E. degree in Computer Science and Technology from Shenzhen University, the M.Tech. degree in Intelligent Systems from National University of Singapore. His research interests include deep learning, multi-modality representation and recommender systems.
\end{IEEEbiography}
\vskip -0.4in
\begin{IEEEbiography}[{\includegraphics[width=1in,height=1.25in,clip,keepaspectratio]{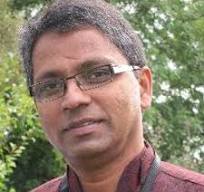}}]{Joemon Jose} is a Professor with the School of Computing Science, University of Glasgow,
Glasgow, and a member of the Information Retrieval
Group. He has published over 300 papers with more
than 10,000 Google Scholar citations, and an H-index
of 51. He leads the Multimedia Information Retrieval
group which investigates research issues related
to the above themes. His research focuses around
the following three themes: social media analytics,
multimodal interaction for information retrieval, and
multimedia mining and search.
\end{IEEEbiography}






\end{document}